\documentclass[12pt]{WileyMSP-template}
\usepackage{textcomp, gensymb}
\usepackage{subfig}
\usepackage{amsmath}
\usepackage{comment}
\usepackage{xcolor}
\usepackage{siunitx}
\usepackage{upgreek}
\usepackage{mathptmx}
\usepackage{cite}
\usepackage{pdfpages}
\usepackage{soul}
\begin{document}
\linespread{1.25}

\title{Exchange bias between van der Waals materials: tilted magnetic states and field-free spin-orbit-torque switching}
\maketitle
\vspace{0.5cm}

\author{Thow Min Jerald Cham\textsuperscript{1$\ddagger$}, Reiley J. Dorrian\textsuperscript{1}, Xiyue S. Zhang\textsuperscript{1}, Avalon H. Dismukes\textsuperscript{2}, Daniel G. Chica\textsuperscript{2}, Andrew F. May\textsuperscript{3}, Xavier Roy\textsuperscript{2}, David A. Muller\textsuperscript{1,4},  Daniel C. Ralph\textsuperscript{1,4*}, Yunqiu Kelly Luo\textsuperscript{1,4,5$\ddagger$*}}

\vspace{0.5cm}
\textsuperscript{1} Cornell University, Ithaca, NY 14850, USA\\
\textsuperscript{2} Department of Chemistry, Columbia University, New York, NY 10027, USA\\
\textsuperscript{3} Materials Science and Technology Division, Oak Ridge National Laboratory, Oak Ridge, TN 37831, USA\\
\textsuperscript{4} Kavli Institute at Cornell, Ithaca, NY 14853, USA\\
\textsuperscript{5} Department of Physics and Astronomy, University of Southern California, Los Angeles, CA 90089, USA\\
\textsuperscript{*}Email: dcr14@cornell.edu\\
\textsuperscript{*}Email: yl664@cornell.edu, kelly.y.luo@usc.edu\\
\textit{\textsuperscript{$\ddagger$} T.M.J.C and Y.K.L contributed equally to this paper}

\vspace{0.5cm}
\textbf{
Magnetic van der Waals heterostructures provide a unique platform to study magnetism and spintronics device concepts in the two-dimensional limit. Here, we report studies of  exchange bias from the van der Waals antiferromagnet CrSBr acting on the van der Waals ferromagnet Fe$_\text{3}$GeTe$_\text{2}$ (FGT). The orientation of the exchange bias is along the in-plane easy axis of CrSBr, perpendicular to the out-of-plane anisotropy of the FGT, inducing a strongly tilted magnetic configuration in the FGT.  Furthermore, the in-plane exchange bias provides sufficient symmetry breaking to allow deterministic spin-orbit torque switching of the FGT in CrSBr/FGT/Pt samples at zero applied magnetic field.  A minimum thickness of the CrSBr greater than 10 nm is needed to provide a non-zero exchange bias at 30 K.} 

\vspace{0.5 cm}
The discovery of spin ordering persisting to the monolayer limit in vdW magnets provides a new platform for exploring magnetic phenomena that are tunable through electric gating\textsuperscript{\cite{xing2017electric, jiang2018controlling, chen2018role, zheng2020gate, tan2021gate}}, optical control\textsuperscript{\cite{Wang2022,Bae2022}}, layer twisting\textsuperscript{\cite{sivadas2018stacking, tong2019magnetic, song2021direct, Xu2022}}, and strain or pressure\textsuperscript{\cite{webster2018strain, song2019switching, li2019pressure, wang2020strain, cenker2022reversible, diederich2023tunable}}. vdW magnets are easily integrated into heterostructures with other vdW materials for exploring interfacial effects and new device function\-alities,\textsuperscript{\cite{seyler2018valley, zhong2020layer, huang2020emergent, choi2023asymmetric}} and can have magnetic structures distinct from conventional bulk magnets because the interlayer exchange is much weaker than the intralayer exchange\textsuperscript{\cite{fang2018large}}. These characteristics can be leveraged to realize non-trivial spin ordering such as coexisting antiferromagnetic and ferromagnetic order\textsuperscript{\cite{Xu2022}} or skyrmions\textsuperscript{\cite{ding2019observation, wu2022van, casas2023coexistence}}. Exchange-bias interactions\textsuperscript{\cite{meiklejohn1956new}} between vdW antiferromagnets (AFs) and ferromagnets (FMs) also have the potential to be particularly interesting because the exchange interaction at the AF/FM interface may be comparable to the interlayer interactions within each material, producing the possibility of spatially non-uniform spin configurations not readily achievable with conventional magnetic materials. However, experiments probing exchange bias in vdW materials have thus far only considered exchange bias parallel to the anisotropy axis of the FM, visible as shifts of hysteresis curves as a function of applied magnetic field along that axis\textsuperscript{\cite{Zhu2020, Zhang2020_FePS, Dai2021, Wu2022, Zhang2022, Jo2022}} -- shifts which have been interpreted in the framework of an effectively macrospin response. Exchange interactions perpendicular to the FM's anisotropy axis in vdW heterostructures and the resulting spatially non-uniform magnetic configurations remain unexplored.

\vspace{0.5cm}
We study the interaction between ferromagnetic Fe$_3$GeTe$_2$ (FGT)\textsuperscript{\cite{Tan2018}} with perpendicular magnetic anisotropy (PMA) and antiferromagnetic CrSBr\textsuperscript{\cite{Telford2020, Lee2021, Telford2022}} with in-plane easy-axis anisotropy, and find that the interaction induces an {\em in-plane} exchange bias on the FGT. We determine an effective exchange field as strong as B$_\text{ext} \approx 0.15$ T at 10 K, comparable to the interlayer exchange within vdW magnets. With zero applied field and no applied current, the in-plane exchange field produces a strong tilting of the FGT magnetization. Since the exchange bias is an interface interaction and the anisotropy in FGT arises from a bulk mechanism, we conclude that the tilting is non-uniform through the thickness of FGT (illustrated schematically in Fig.~1a).  Furthermore, in CrSBr/FGT/Pt samples we observe current-driven switching of FGT without any applied magnetic field using spin-orbit torque (SOT) from the Pt layer, indicating that the in-plane exchange bias provides the necessary symmetry-breaking to make the SOT switching deterministic for a magnet with PMA.

\vspace{0.5cm}
FGT is a metallic vdW ferromagnet with strong intra-layer and weaker inter-layer ferromagnetic coupling, and a Curie temperature (T$_\text{C}$) of 160-190 K.\textsuperscript{\cite{Tan2018}} CrSBr is a semiconducting vdW antiferromagnet with ferromagnetic intra-layer coupling and antiferromagnetic inter-layer coupling, with antiferromagnetic order below a N\'eel temperature (T$_\text{N}$) of 132 K\textsuperscript{\cite{Telford2020}}. Above this temperature, CrSBr also has an intermediate ferromagnetic phase up to $\approx$160 K\textsuperscript{\cite{Lee2021}}. Prior measurements on CrSBr indicate triaxial magnetic anisotropy with an in-plane easy axis along the crystallographic b axis (Fig.~1)\textsuperscript{\cite{Lee2021, Yang2021, Telford2022, Cenker2022, Ye2022, Bae2022, Cham2022, Diederich2023}}. CrSBr flakes typically exfoliate into thin strips along the crystallographic a axis. This allows us to orient the N\'eel vector of the CrSBr flake in any desired in-plane direction.
 
 \vspace{0.5cm}
 We first probe the effect of interfacial exchange coupling from CrSBr on the magnetic order of FGT using anomalous-Hall resistance measurements on CrSBr/FGT/Pt samples while sweeping an out-of-plane magnetic field. The much higher resistance of CrSBr relative to FGT means that the overall Hall resistance can be used as a read-out of just the FGT magnetization\textsuperscript{\cite{Wang2019, Telford2022}}. Fig.~2a shows the out-of-plane field sweeps (10-200 K) measured beginning with FGT saturated out-of-plane, showing hysteresis up to 180 K near the FGT T$_\text{C}$. We do not observe any horizontal shifts indicative of an out-of-plane exchange bias on the FGT (S.I.III). Notably, the hysteresis is qualitatively different above and below T$_\text{N}$ (Fig.~2b). At 140 K, the loop is close to square with a remnant magnetization at 0 T close to the saturation value, characteristic of FGT with uniform out-of-plane magnetization\textsuperscript{\cite{Tan2018, Wang2019}}. FGT/Pt samples without CrSBr show similar close-to-square behavior from 160 K to 10 K (S.I. Fig.~S1). However, for the CrSBr/FGT/Pt sample at 10 K the hysteresis loop is far from square -- there is a gradual decrease in Hall resistance as the field is swept through 0 T, indicating a reduction in the out-of-plane magnetization.  If the out-of-plane field is swept from saturation to zero and then reversed (i.e., a minor loop), the Hall resistance is not hysteretic (S.I. Fig.~S1e,f). This demonstrates that in the zero-field state the FGT magnetization has not broken up into a mix of lateral domains with positive and negative out-of-plane components, but rather the average magnetization in FGT is simply tilted away from the out-of-plane direction.
 
 \vspace{0.5cm}
 We have also studied how magnetic fields applied in the sample plane affect the magnetic configuration of the CrSBr/FGT/Pt samples. After saturating the FGT out-of-plane, we reduce the applied magnetic field to zero and sweep an in-plane field back and forth along the magnetic easy-axis axis of the CrSBr. At 30 K, well below T$_\text{N}$, in-plane sweeps up to 0.8 T give an overall approximately parabolic dependence that is non-hysteretic (Fig.~2c). We do not observe a directional exchange bias in these curves, in that to experimental accuracy they are symmetric about 0 T. We suggest that this is because the initial field used to saturate the FGT out of plane (0.2 T) can be comparable to the exchange and anisotropy fields of the CrSBr\cite{Cham2022} so that the CrSBr moments are also intially forced out-of-plane, and when the field is reduced to 0 T the CrSBr likely relaxes into a multidomain state with a mix of domains having interface layers with opposite orientations along the in-plane easy axis of the CrSBr. Lateral domains within the FGT layer are created only when the in-plane field is swept beyond about 0.8 T, as is evident from  the onset of hysteresis as a function of magnetic field (Fig.~2e). When the temperature is increased beyond T$_\text{N}$ to 140 K, the maximum Hall resistance amplitude for both in-plane and out-of-plane sweeps match, as expected for a purely out-of-plane saturated magnetic state (Fig.~2e). The fractional reduction in the zero-field Hall amplitude compared to the saturated value shows a very large average tilt angle away from out-of-plane for the FGT magnetization at low temperature ($>$50$^\circ$, Fig.~2f). The reduction in the zero-field Hall signal reduces to zero beyond T$_\text{N} \approx$ 132 K, confirming that the tilted state in the FGT is associated with the antiferromagnetism of CrSBr.
 
 \vspace{0.5 cm}
 We have also investigated whether, despite challenges in controlling the domain structure within the antiferromagnetic CrSBr, the exchange interaction from the CrSBr can provide a sufficient symmetry-breaking field to allow for deterministic SOT switching in CrSBr/FGT/Pt heterostructures. In general, to achieve deterministic switching of a magnetic layer with PMA using SOT from a high-symmetry material like Pt requires an external symmetry-breaking field,\textsuperscript{\cite{Liu2012}} which can be accomplished, e.g., with an applied magnetic field in the sample plane parallel to the direction of applied current (B$_\text{ext}$)\textsuperscript{\cite{Miron2011, Liu2012_Pt}}, in-plane exchange bias from an adjacent antiferromagnet or ferromagnet layer\textsuperscript{\cite{Oh2016, Fukami2016, lau2016spin, VanDenBrink2016}}, or other effective fields\textsuperscript{\cite{Wu2022_mater}}. We perform pulsed-current measurements with different fixed B$_\text{ext}$, and after each pulse we measure the Hall voltage near zero current. Before each pulse-current switching sequence, we initialize the magnetization of the FGT with an out-of-plane field just above the coercivity. This field is then ramped back to zero, before the respective in-plane fields were applied.
 
 \vspace{0.5cm}
 Figures 3a-c show the resulting switching loops for the device with $\hat{\text{N}} \parallel$ I at 30 K. We observe deterministic switching (Fig.~3b) for B$_\text{ext}$ = 0 T, with the same switching chirality as when B$_\text{ext}$ = -0.1 T (Fig.~3a). When B$_\text{ext}$ = 0.1 T, we see a quenching of the hysteresis. In comparison, in the device for which $\hat{\text{N}}$ $\perp$ I, there is negligible hysteresis at 0 T (Fig.~3d) and the chirality of the magnetization reversal is opposite for $\pm$0.1 T (Fig.~3e,f). When the temperature is raised above T$_\text{N}$ to 170 K in the device with $\hat{\text{N}} \parallel$ I, we see no switching at 0 T (Fig.~4e) and opposite switching chiralities at $\pm$0.05 T (Fig.~4e,f). These findings indicate that a net exchange bias is induced parallel to the N\'eel vector of CrSBr when the temperature is lowered below T$_\text{N}$. This in-plane uniaxial exchange bias appears without the application of any in-plane magnetic field during the cooling process, and is unchanged in sign upon field cooling in fields up to 8 T (S.I.VII). 
 
\vspace{0.5cm}
 We note that the switching loops show incomplete magnetization reversal. Even before the application of any current pulses, the samples begin in a state with a Hall signal less than the saturated value as described above, and then the amplitude of the current-driven hysteresis is smaller still. The existence of incomplete reversal is similar to previous reports of SOT switching of FGT/Pt heterostructures without CrSBr,\textsuperscript{\cite{Wang2019, Alghamdi2019, Zhang2021}} where this behaviour was attributed to a thermally-induced multidomain state at large currents.\textsuperscript{\cite{Wang2019}} We describe micromagnetic simulations that support this interpretation in S.I.X. 

\vspace{0.5cm}
 We further investigate the current-driven magnetization switching amplitude $\Delta$R (Fig.~4a inset) at different values of B$_\text{ext}$. We show results for two devices: $\hat{\text{N}}\parallel$ I and $\hat{\text{N}}\perp$ I at 170 K and 30 K (Fig.~4a-d). At 170 K, $\Delta$R of both devices follow approximately an anti-symmetric Gaussian lineshape: $\Delta$R $= 0$ at B$_\text{ext} = 0$,  and then $|\Delta$R$|$ increases as $|$B$_\text{ext}|$ increases, reaching a peak before decreasing to zero at large B$_\text{ext}$. The switching chirality, or sign of $\Delta$R, depends on the sign of B$_\text{ext}$, as expected for SOT mediated switching\textsuperscript{\cite{Liu2012}}. At 30 K, we see a similar anti-symmetric Gaussian field dependence for the $\hat{\text{N}}$ $\perp$ I device (Fig.~4c), indicating the absence of exchange bias along the current direction. However, a net exchange bias is evident for the $\hat{\text{N}}$ $\parallel$ I device (Fig.~4c), in that the overall curve is shifted along the field axis, such that $\Delta$ R $< 0$ at B$_\text{ext} = 0$ and a non-zero positive B$_\text{ext}$ offseting the exchange-bias is required to drive $\Delta$R to zero. Fig.~4e shows the extracted effective exchange bias (B$_\text{EB}$) from linear fits to $\Delta$R in the low field limit as a function of temperature. For the $\hat{\text{N}} \parallel$ I device, B$_\text{EB}$  increases gradually as the temperature is lowered below about 170 K, up to $\approx$ 0.15 T at 10 K. However when $\hat{\text{N}}$ $\perp$ I, B$_\text{EB}$ is negligible. The estimated exchange bias for $\hat{\text{N}} \parallel$ I is within about a factor of two of the CrSBr interlayer exchange determined previously from antiferromagnetic resonance measurements,\textsuperscript{\cite{Cham2022}} with similar temperature dependence (S.I.IV).

 \vspace{0.5cm}
 Previous measurements of out-of-plane exchange bias in FGT/antiferromagnet heterostructures showed that the strength of the exchange bias depends on the antiferromagnet thickness\textsuperscript{\cite{Zhu2020, Jo2022}}, with a critical thickness needed for a non-zero exchange bias effect. We performed pulsed-current hysteresis measurements to estimate the exchange bias for four different devices with $\hat{\text{N}} \parallel$ I,  with CrSBr thicknesses ranging from 10.5 nm to 47 nm, along with a Pt/FGT reference device with no CrSBr, all with similar FGT thicknesses (9-12 nm). We see that the Pt/FGT reference sample and the thinnest (t$_\text{CrSBr}$ $\approx$ 10 nm) sample show negligible B$_\text{EB}$, while the samples with t$_\text{CrSBr}$ from 30 to 47 nm show non-zero exchange bias (Fig.~4f and S.I.VI). The presence of a critical CrSBr thickness for a non-zero B$_\text{EB}$ further indicates that this effect arises from the exchange interaction between FGT and CrSBr.

\vspace{0.5cm}
In conclusion, we report measurements of an in-plane exchange bias from CrSBr acting perpendicular to the out-of-plane anisotropy of FGT, in a direction parallel to the in-plane anisotropy axis of CrSBr. This exchange field results in a strongly-tilted magnetic configuration within the FGT, and can serve as an in-plane symmetry-breaking field that enables field-free deterministic switching driven by SOT in CrSBr/FGT/Pt devices. Although the CrSBr in our samples is likely in a multidomain state, we can make a rough estimate of the exchange bias strength from the  external magnetic field required to cancel the exchange field and eliminate the deterministic switching. We estimate values as large as 0.15 T at low temperature, decreasing gradually with increasing temperature up to the T$_\text{N}$ of CrSBr. A CrSBr thickness greater than $\approx$10 nm is required to provide exchange bias for switching at 30 K. This work opens possibilities for exploiting unique characteristics of vdW magnets and heterostructures to enable new functionality in spintronics.

\section*{Methods}
CrSBr single crystals were synthesized using a modified chemical vapor transport approach originally adapted from Beck\cite{Beck1990}. Two methods were used for the synthesis of CrSBr crystals which produced crystals of identical structure, composition, and magnetic properties. Crystals of CrSBr from Method 1 were used in all devices except for the device containing the 47 nm thick CrSBr flake which used CrSBr crystals from Method 2. Method 1 used disulfur dibromide and chromium metal as reagents which were added together in a 7:13 molar ratio to a fused silica tube approximately 35 cm in length. This tube was sealed under vacuum and placed in a three-zone tube furnace. The tube was heated in a temperature gradient (950 $\degree$C to 850 $\degree$C) for 120 hours. Method 2 used chromium, sulfur, and chromium tribromide as reagents in a slightly off stoichiometric ratio which were sealed in a fused silica tube of 20 cm. The tube was subjected to a modified heating profile using a two-zone tube furnace with a temperature gradient of 950 $\degree$C to 850 $\degree$C. Further details of the synthesis can be found in \cite{Telford2020} and \cite{Scheie2022} for Method 1 and 2, respectively.

\vspace{0.5cm}
Bulk Fe$_\text{3-x}$GeTe$_\text{2}$ crystals were synthesized from a self-flux using an initial composition of Fe$_\text{6}$GeTe$_\text{9}$, which was homogenized above 1150 $\degree$C then cooled to 750 $\degree$C and the crystals were isolated from the flux by centrifugation. This approach yields FGT with a bulk Curie temperature above 200 K as described in ref.\cite{MayYanMcGuire2020,Drachuck2018}.

\vspace{0.5cm}
Pre-patterned Hall bars were prepared using standard photolithography processes and e-beam evaporation of 10 nm of Pt, patterned into devices with widths of 20 $\mu$m and lengths of 60 $\mu$m, with Hall leads 4 $\mu$m wide and 20 $\mu$m long. A liftoff procedure was utilized to avoid contact of the top surface of the Pt layer with photoresist. The prepared Pt bars were placed into an Ar glove box with H$_2$O and O$_2$ levels $<$0.5 ppm where further processing was done. First the Hall bars were heated on a hotplate at 180$\degree$C to remove any residual adsorbed water. CrSBr and Fe$_\text{3}$GeTe$_\text{2}$ were then exfoliated onto high-resistivity silicon/silicon dioxide (280 nm) wafers using the scotch-tape method. Flakes of appropriate thicknesses were selected using an optical microscope equipped with a differential interference contrast prism for enhancing the optical contrast of steps in the flakes. The selected flakes were then mechanically transferred\textsuperscript{\cite{ponomarenko2011, geim2013}} using stamps made from polypropylene carbonate (PPC) and polycarbonate (PC) onto the Hall bars. The FGT flakes typically have a size of around 10 $\mu$m $\times$ 10 $\mu$m and CrSBr flakes a size of around 15 $\mu$m $\times$ 30 $\mu$m. For each device, a CrSBr flake larger than the FGT flake was used such that FGT was covered in its entirety, but the FGT flake did not extend across the entire width of the Hall bar.  After completion of the transfer, polymer residue was removed in chlorofoam, and the devices rinsed in acetone and then IPA before measurements. To prevent degradation, the devices were capped with an hBN layer except the two with the thickest CrSBr layers (42 nm and 47 nm), and were also stored in a glovebox between measurements. We use atomic force microscopy to quantify the thicknesses and uniformity of the layers in each device. We verify that the CrSBr layers have no monolayer steps, and that the FGT layers are uniform except for some steps near the edges of the flakes.

\vspace{0.5cm}
Anomalous Hall effect and spin-orbit torque switching measurements were performed in a Quantum Design EverCool PPMS with a maximum magnetic field of 9 T. The sample was moved between the field-in-plane and field-out-of-plane configurations using a rotator. The sample was aligned on the holder such that the in-plane field was parallel to the current direction. Current pulses were applied using a Keithley 2400 sourcemeter with a pulse length of 50 microseconds, while the Hall voltage was measured using a Signal recovery 7265 lock-in amplifier with an output frequency of 1117.17 Hz.

\vspace{0.5cm}
\textbf{Acknowledgements} \par
We thank Rakshit Jain, Patrick Kn\"{u}ppel, Steve Kriske, Liguo Ma, Aaron Windsor, and Jiacheng Zhu for experimental assistance, and inspiring discussions with M\"{a}rta Tschudin, Arnab Bose, John Cenker, Vishakha Gupta, Shengwei Jiang,  Kaifei Kang,  Kihong Lee, Kin Fai Mak, and Jie Shan. The research at Cornell was supported by the AFOSR/MURI project 2DMagic (FA9550-19-1-0390) and the US National Science Foundation through the Cornell Center for Materials Research (DMR-1719875). T.M.J.C. was supported by the Singapore Agency for Science, Technology, and Research, and Y.K.L. acknowledges a Cornell Presidential Postdoctoral Fellowship. The work utilized the shared facilities of the Cornell Center for Materials Research and the Cornell NanoScale Facility, a member of the National Nanotechnology Coordinated Infrastructure (supported by the NSF via grant NNCI-2025233), and it benefited from instrumentation support by the Kavli Institute at Cornell. Synthesis of the CrSBr crystals was supported as part of Programmable Quantum Materials, an Energy Frontier Research Center funded by the U.S. Department of Energy (DOE), Office of Science, Basic Energy Sciences (BES), under award DE-SC0019443, and the Columbia MRSEC on Precision-Assembled Quantum Materials (PAQM) under award number DMR-2011738. FGT crystal growth and characterization (AFM) was supported by the U. S. Department of Energy, Office of Science, Basic Energy Sciences, Materials Sciences and Engineering Division. Electron microscopy was supported by the PARADIM NSF Materials Innovation Platform (DMR-2039380). 

\vspace{0.5cm}
\textbf{Author Contributions} \par
T.M.J.C. and Y.K.L. devised the experiment, fabricated the vdW heterostructures and performed the measurements. T.M.J.C performed the data analysis with assistance from Y.K.L.. R.J.D. performed the micromagnetic simulations with assistance from T.M.J.C. and Y.K.L.. X.S.Z. performed the STEM imaging, supervised by D.A.M.  A.H.D. and D.G.C. synthesized the CrSBr crystals supervised by X.R.. A.F.M. synthesized the FGT crystals. D.C.R. provided oversight and advice. T.M.J.C, Y.K.L., and D.C.R. wrote the manuscript. All authors discussed the results and the content of the manuscript.
All authors discussed the results and the content of the manuscript.

\vspace{0.5cm}
\textbf{Competing Interests} \par
The authors declare no competing interests.

\bibliographystyle{MSP}
\bibliography{main}

\newpage
\textbf{Fig 1. Device schematic and crystal structure} Schematic of a CrSBr/FGT heterostructure dry transferred onto a Pt channel for spin-orbit torque pulse current switching measurements. The average out-of-plane magnetization of the FGT layers (orange arrows) is inferred from the Hall voltage readout transverse to the Pt channel. The magnetic moments of the CrSBr layers (blue arrows) are aligned ferromagnetically within each layer and antiferromagnetically between adjacent layers. (b) Top-view optical image of a CrSBr(30 nm)/FGT(9 nm)/Pt (10 nm) device with the b crystal axis of the CrSBr layer oriented parallel to the current. (c) High angle annular dark field (HAADF) STEM cross-sectional image of the vdW interface of a CrSBr/FGT heterostructure. As seen from the overlaid atoms, we confirm the orientation of the b axis of CrSBr layer, which in this case was aligned parallel to the Pt channel ($\hat{N}$ $\parallel$ I).

\vspace{0.5cm}
\textbf{Fig 2. Anomalous Hall resistance of CrSBr/FGT/Pt samples with out-of-plane and in-plane field sweeps} (a) Anomalous Hall effect (AHE) resistance hysteresis loops as a function of applied magnetic field swept in the out-of-plane direction, for temperatures between 10 K to 200 K, showing a Curie temperature for the FGT close to 180 K. Additional steps in the hysteresis loops may come from thickness variation in the FGT layer or domain formation. (b) Close-up of the AHE hysteresis loops at 10 K and 140 K. (c,d) Anomalous Hall resistance comparing out-of-plane field sweeps to in-plane sweeps along the easy-anisotropy b axis of the CrSBr at (c) 10 K and (d) above the CrSBr N\'eel temperature at 140 K. R$_\text{max}$ indicates the Hall resistance when the FGT magnetization is fully-saturated out of plane and R$_0$ indicates the value at zero applied magnetic field. (e) Hall resistance hysteresis loops as a function of in-plane applied magnetic field for different maximum values of the in-plane field. (f) Reduction of the Hall resistance as a function of temperature extracted from the difference between R$_\text{max}$ and R$_{0}$ for samples with $\hat{N}$ $\parallel$ I (30 nm CrSBr/9 nm FGT/10 nm Pt) and $\hat{N}$ $\perp$ I (37 nm CrSBr/12 nm FGT/10 nm Pt). 

\vspace{0.5cm}
\textbf{Fig 3. Pulsed-current-switching hysteresis loops of CrSBr/FGT/Pt samples.} For each loop, we start the pulsed-current sequence from 0 pulsed current amplitude, go up to the maximum positive current, down to the minimum negative current, and finally back up to the maximum positive current (black arrows). (a-c) Pulsed-current-switching hysteresis loops at 30 K for a sample (30 nm CrSBr/9 nm FGT/10 nm Pt) with the b axis of CrSBr aligned along the Pt channel, such that $\hat{N}$ $\parallel$ I. In this configuration, the uniaxial exchange bias field enables deterministic spin-orbit-torque switching within the FGT layer at 0 T (panel (b)). (c) With a positive field of 0.1 T, the magnetization reversal hysteresis is quenched. (d-f) Pulsed-current switching hysteresis loops at 30 K for a sample (37 nm CrSBr/12 nm FGT/10 nm Pt) with the a axis of CrSBr aligned along the Pt channel, such that $\hat{N}$ $\perp$ I. In contrast with the first configuration, we observe no deterministic spin-orbit-torque-driven magnetization reversal at 0 T (panel (e)). With fields of (d) -0.1 T and (f) +0.1 T, we observe hysteresis loops of opposite chiralities, as expected for a spin-orbit-torque-driven magnetization reversal process. (g-i) Pulsed-current-switching hysteresis loops at 170 K, for the device with $\hat{N}$ $\parallel$ I. Unlike the behavior observed at 30 K, above the N\'eel temperature of CrSBr no deterministic field-free spin-orbit torque driven magnetization reversal is observed (panel (h)) and the hysteresis loops for external fields of opposite signs  have opposite chiralities (panels (g,i)).

\vspace{0.5cm}
\textbf{Fig 4. Magnetization reversal amplitude of CrSBr/FGT/Pt samples as a function of external fields B$_\text{ext}$ applied parallel to the current axis.} (a-d) $\Delta$R vs.\ B$_\text{ext}$ plots for devices with both $\hat{N}$ parallel (30 nm CrSBr/9 nm FGT/10 nm Pt) and perpendicular (37 nm CrSBr/12 nm FGT/10 nm Pt) to I  at 30 K and 170 K. We define the magnetization reversal amplitude $\Delta$R as shown in the inset of panel (a) as the difference in Hall resistance between the state that results from spin-orbit switching at negative current (blue) and the state that results from switching at positive current (red).  $\Delta$R values are taken from the averaged difference while uncertainties are calculated from the standard deviation of R$_{xy}$ values in the upper and lower states. At 170 K, both devices show a similar external field dependence, with $\Delta$R $\approx$ 0 at 0 T and $\Delta$R of opposite signs for opposite fields (panels (a) and (b)). At 30 K, while the ($\hat{N}$ $\perp$ I) device exhibits the same field dependence as the 170 K plot with $\Delta$R = 0 at B$_\text{ext}$ = 0 T, the ($\hat{N}$ $\parallel$ I) device shows a distinctly different field dependence $\Delta$R $<$ 0 at B$_\text{ext}$ = 0 T. Linear fits to the low field regime are used to extract the external field strength at which the magnetization reversal is quenched (dashed lines). (e) Effective exchange-bias field strength as a function of temperature for both devices. For the ($\hat{N}$ $\parallel$ I) device (red points) the exchange bias increases as T is decreased from 170 K to 10 K. In contrast, the ($\hat{N}$ $\perp$ I) device does not show any significant exchange bias parallel to the current direction. (f) Effective exchange-bias field strength as a function of CrSBr thickness, indicating a critical CrSBr thickness for a non-zero B$_\text{EB}$. The four additional devices have layer thicknesses FGT(10 nm)/Pt(10 nm) with no CrSBr, CrSBr(10.5 nm)/FGT(9 nm)/Pt(10 nm), CrSBr(42 nm)/FGT(11.6 nm)/Pt(10 nm) and CrSBr(47 nm)/FGT(9.6 nm/Pt(10 nm)).

\begin{figure}[htpb]
    \centering
    \includegraphics[width=1.0\textwidth]{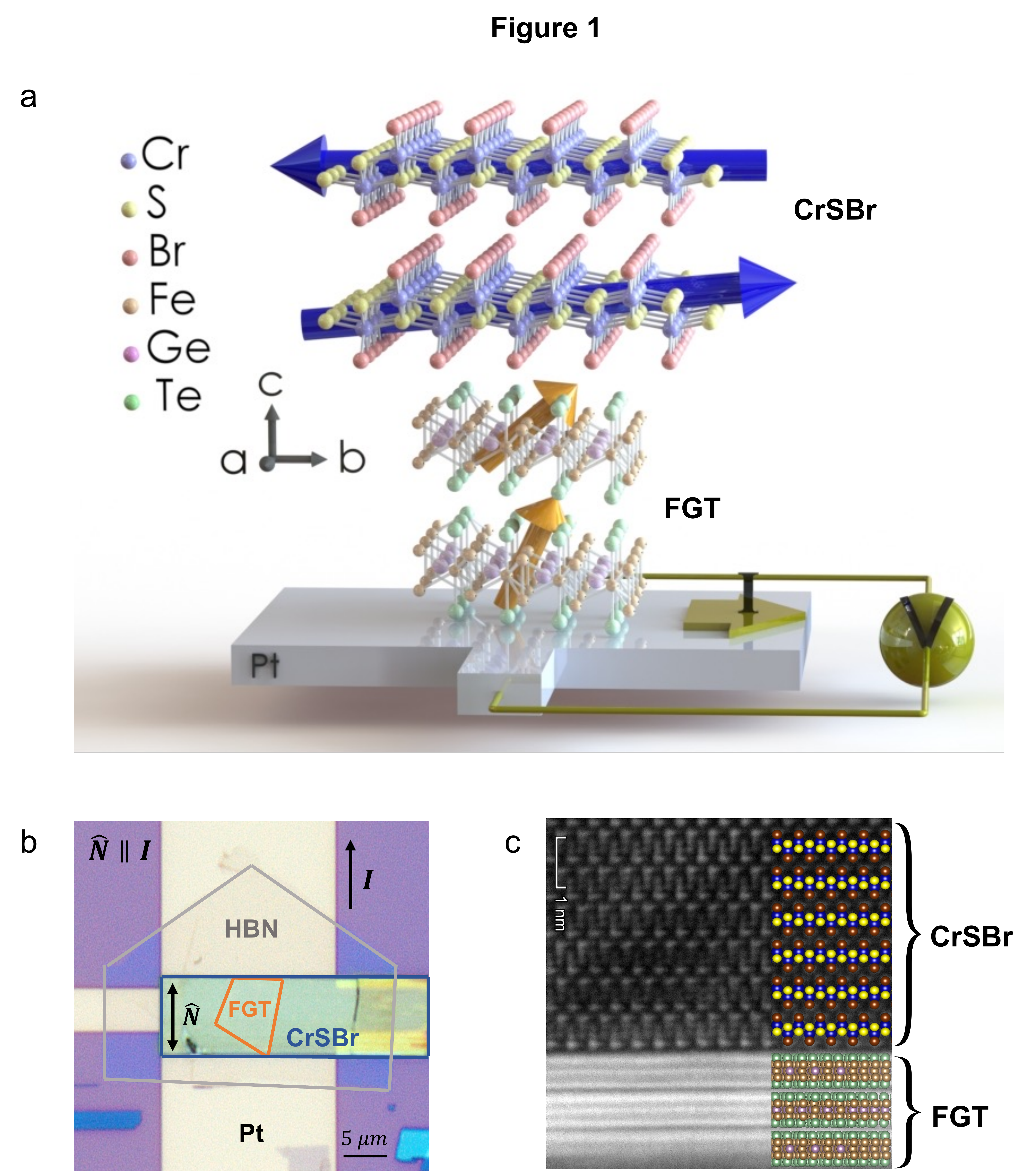}
    \label{Figure1}
\end{figure}

\begin{figure}[htpb]
    \centering
    \includegraphics[width=0.9\textwidth]{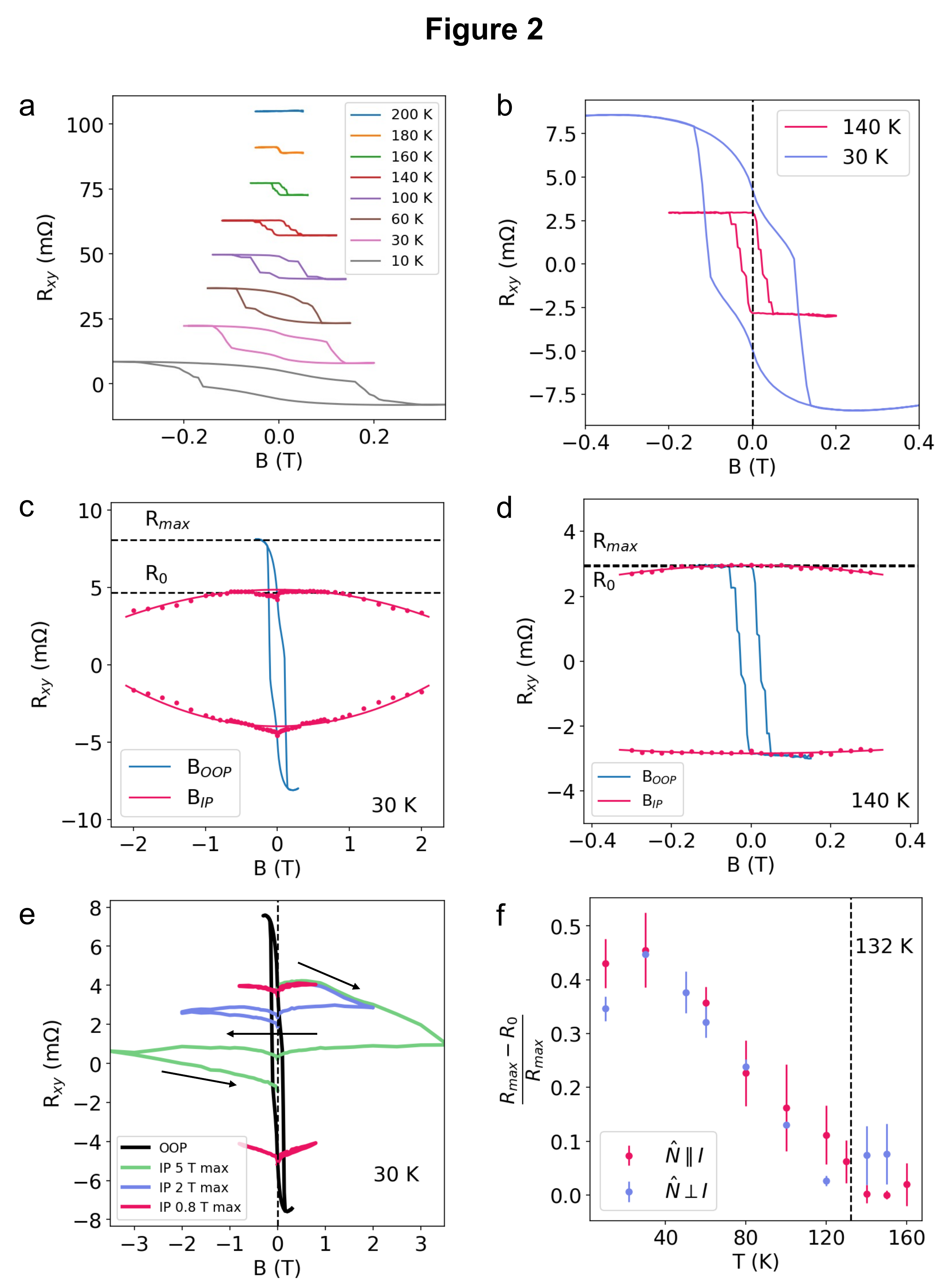}
    \label{Figure2}
\end{figure}

\begin{figure}[htpb]
    \centering
    \includegraphics[width=0.9\textwidth]{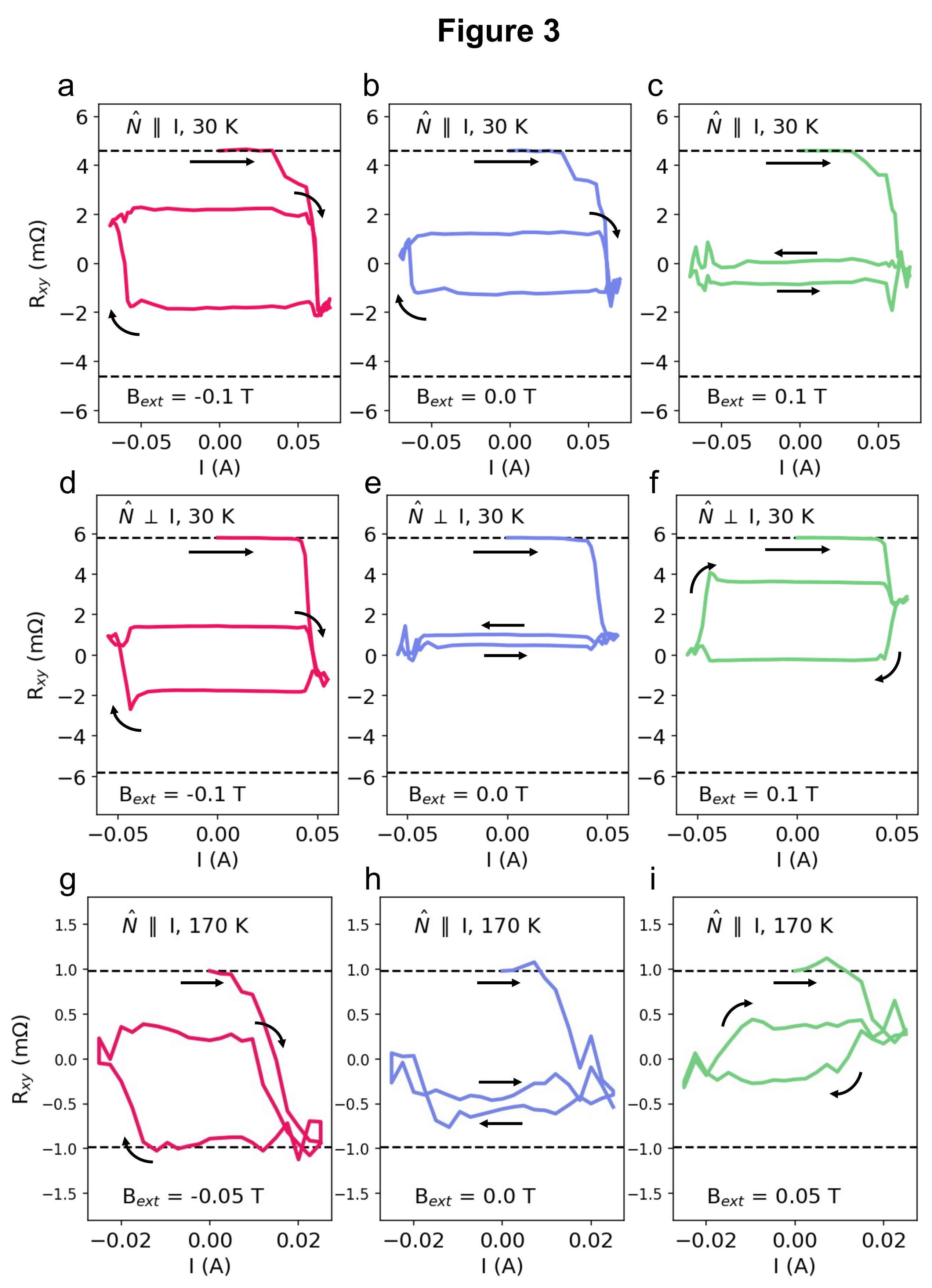}
    \label{Figure3}
\end{figure}

\begin{figure}[htpb]
    \centering
    \includegraphics[width=1.0\textwidth]{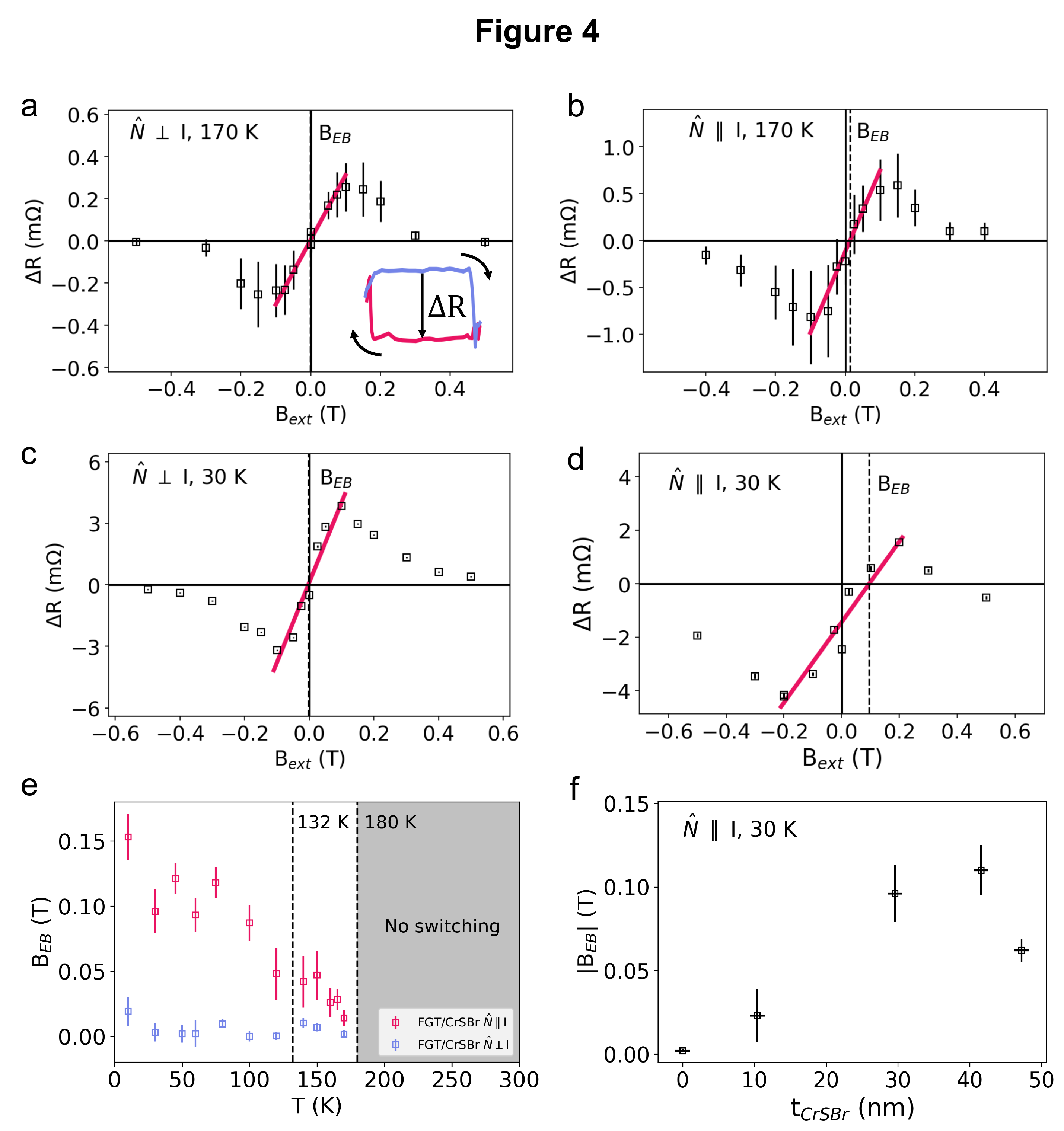}
    \label{Figure4}
\end{figure}

\newpage
\begin{center}
    \textbf{\LARGE{Supplementary Information}}\\
    \vspace{5 mm}
     \textbf{\LARGE{Exchange bias between van der Waals materials: tilted
magnetic states and field-free spin-orbit-torque switching}}\\

\noindent\normalsize{Thow Min Jerald Cham\textsuperscript{1$\dagger$}, Reiley J. Dorrian\textsuperscript{1$\dagger$}, Xiyue, S. Zhang\textsuperscript{1}, Avalon H. Dismukes\textsuperscript{2}, Daniel G. Chica\textsuperscript{2}, Andrew F. May\textsuperscript{3}, Xavier Roy\textsuperscript{2}, David A. Muller\textsuperscript{1,4}, Daniel C. Ralph\textsuperscript{1,4*},}\\
\noindent\normalsize{Yunqiu Kelly Luo\textsuperscript{1,4,5$\dagger$*}}\\
\vspace{5mm}
\small{\noindent\textit{\textsuperscript{1}Cornell University, Ithaca, NY 14850, USA}}\\
\textit{\textsuperscript{2}Department of Chemistry, Columbia University, New York, NY 10027, USA}\\
\textit{\textsuperscript{3}Materials Science and Technology Division, Oak Ridge National Laboratory, Oak Ridge, TN 37831, USA}\\
\textit{\textsuperscript{4}Kavli Institute at Cornell, Ithaca, NY 14853, USA}\\
\textit{\textsuperscript{5}Department of Physics and Astronomy, University of Southern California,} \\
\textit{Los Angeles, CA 90089, USA}\\
\textit{\textsuperscript{*}Corresponding authors. Email: dcr14@cornell.edu, kelly.y.luo@usc.edu}\\
\textit{\textsuperscript{$\dagger$}These authors contributed equally to this work.}
\end{center}

\tableofcontents
\newpage
\renewcommand{\thesection}{\Roman{section}.} 
\renewcommand{\thesubsection}{\Alph{subsection}.}
\renewcommand{\thefigure}{S\arabic{figure}}
\setcounter{figure}{0}

\section{Temperature dependence of magnetic-field-driven hysteresis loops}
We compare the anomalous Hall effect (AHE) transport measurements of FGT devices with and without a CrSBr layer, for temperatures 10 K to 190 K. As the temperature is increased, we see a decrease in the coercivity of the hysteresis loops, together with a decrease in the Hall resistance, indicative of a reduction in magnetic ordering and out-of-plane anisotropy. We estimate a Curie temperature between 180 K - 190 K from the absence of an AHE signal at 190 K for both devices (Figures S1a $\&$ b), in agreement with measurements of few-layer FGT\textsuperscript{\cite{Tan2018}}. 

\vspace{0.5cm}
For the FGT device, we see a typical square hysteresis loop at both 30 K and 140 K (Figures S1c $\&$ d), with remnant magnetization at 0 T close to the saturation value, characteristic of FGT with uniform out-of-plane magnetization. We measure a close-to-square hysteresis loop for the CrSBr/FGT device at 140 K (Figures S1d), but see a distinctly different loop at 30 K (Figures S1c), below the N\'eel temperature of CrSBr. Here, the remnant magnetization at 0 T is reduced from the saturation value, indicating strong tilting of the magnetizations within the FGT layers
due to an exchange interaction with the CrSBr. The non-hysteretic behavior of the minor loop sweeps from a non-zero field to 0 field and back, (Figures S1e $\&$ f) indicates that in the zero-field state the FGT magnetization has not broken up into a mix of lateral domains with positive and negative out-of-plane components, but rather the average magnetization in the FGT is simply tilted away from the out-of-plane direction.
\vspace{0.5cm}

\begin{figure}[htpb]
    \centering
    \includegraphics[width=1.0\textwidth]{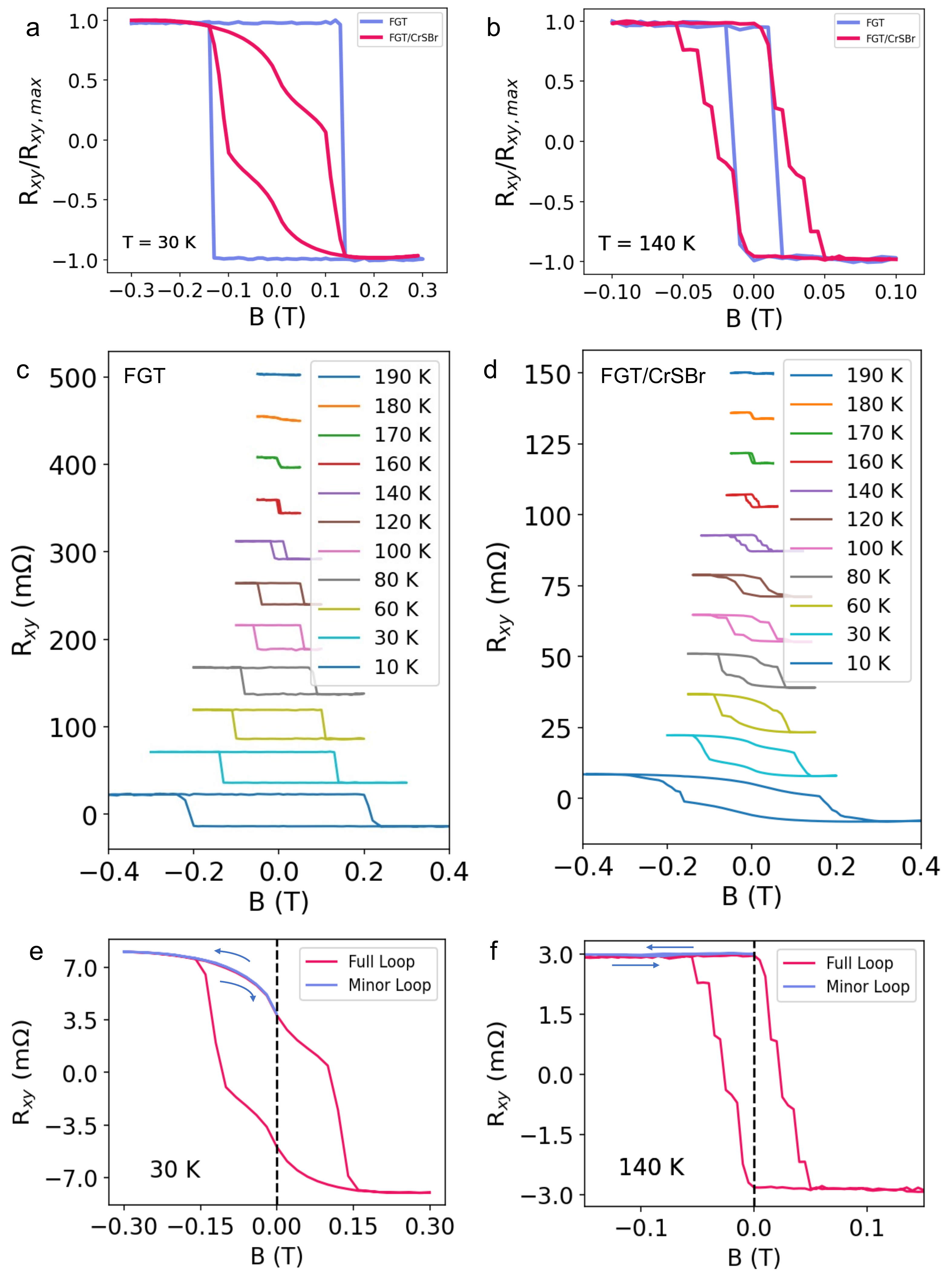}
    \label{SFigure1aa}
\end{figure}

\newpage
\textbf{Fig.~S1. Anomalous Hall resistance hysteresis loops as a function of out-of-plane magnetic field.}  (a,b) AHE loops for (a) FGT(10 nm)/Pt(10 nm) and (b) CrSBr(30 nm)/FGT(9 nm)/Pt(10 nm) from 10 K to 190 K. (c,d) Comparison of AHE loops at (c) 30 K, below the N\'eel temperature of CrSBr and (d) 140 K, close to the N\'eel temperature of CrSBr.  (e,f) Comparison of the full and minor AHE loops for the CrSBr(30 nm)/FGT(9 nm)/Pt(10 nm) device at (e) 30 K and (f) 140 K.

\section{Full Temperature series for measurements of Hall Resistance versus in-plane and out-of-plane magnetic field.}
\begin{figure}[htpb]
    \centering
    \includegraphics[width=0.85\textwidth]{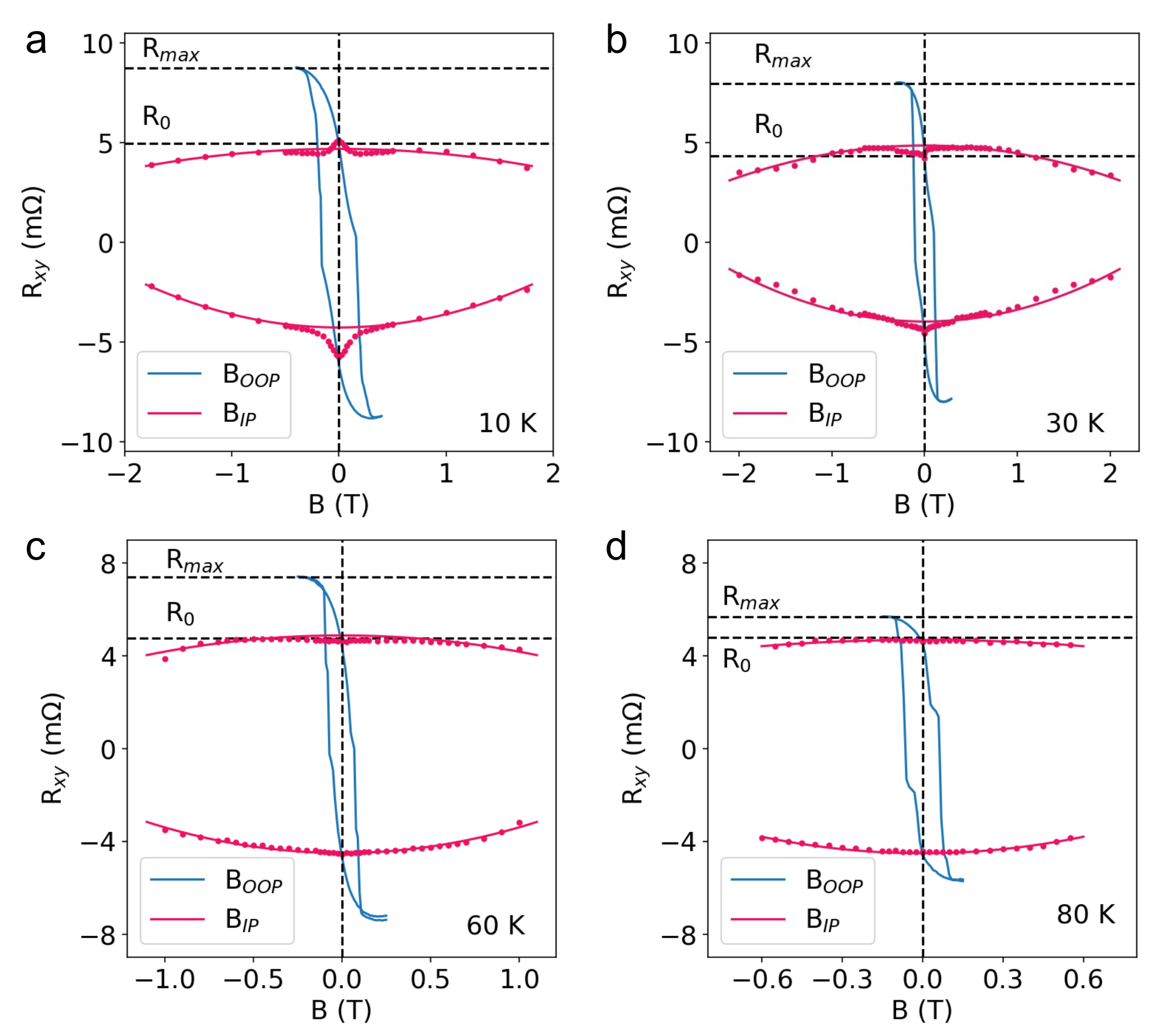}
    \label{SFigure1ab}
\end{figure}

\newpage
\begin{figure}[htpb]
    \centering
    \includegraphics[width=0.79\textwidth]{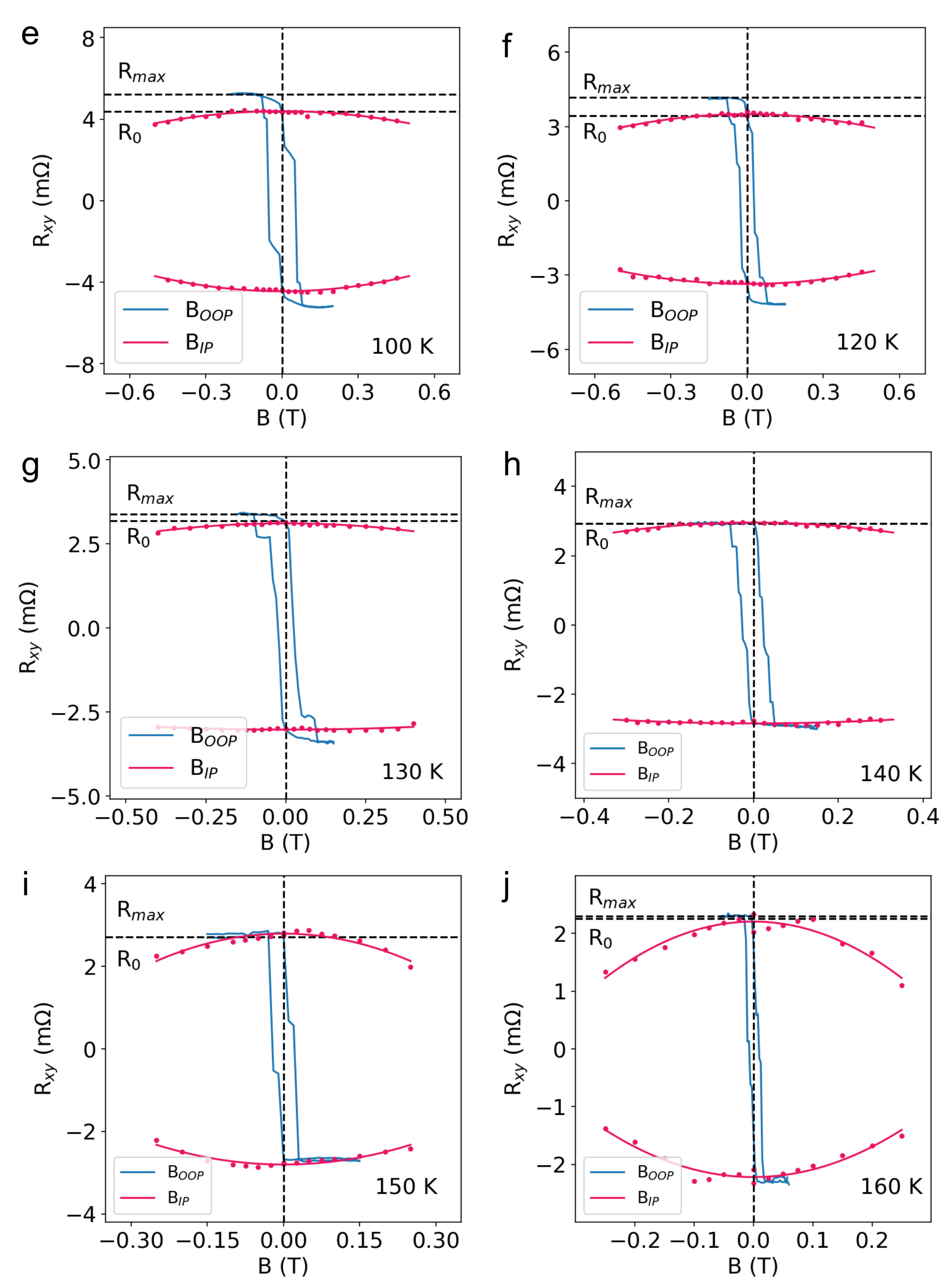}
    \label{SFigure1ab}
\end{figure}

\textbf{Fig.~S2. Temperature dependence of the Hall resistance as a function of out-of-plane and in-plane magnetic field.} Dashed lines indicate the Hall resistances corresponding to the saturated magnetization, R$_\text{max}$, and the remnant value at 0 field, R$_0$.

\newpage
\section{Checking for out-of-plane hysteresis loop shifts}
Previous investigations of the exchange bias effect in some van der Waals heterostructures\textsuperscript{\cite{Zhu2020, Zhang2020_FePS, Dai2021, Wu2022, Zhang2022, Jo2022}} demonstrated out-of-plane exchange bias that results in shifts in the hysteresis curves for magnetic switching as a function of out-of-plane magnetic fields. In the CrSBr/FGT/Pt devices we see negligible shifts in the out-of-plane hysteresis loops (Fig.~S3a). This observation holds for the whole temperature range above and below the N\'eel temperature of CrSBr (Fig.~S3b). This suggests an absence of any out-of-plane exchange bias from CrSBr acting on the FGT.

\begin{figure}[htpb]
    \centering
    \includegraphics[width=1.0\textwidth]{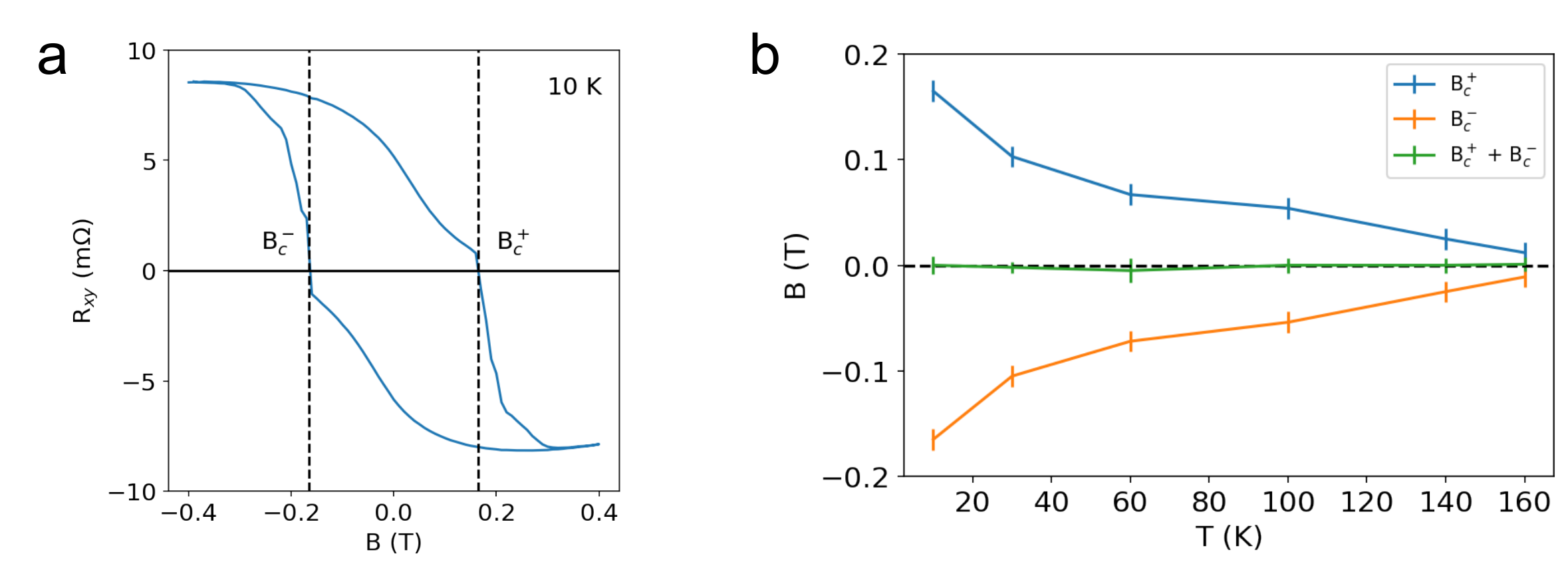}
    \label{SFigure1b}
\end{figure}

\textbf{Fig.~S3. Loop shifts of the out-of-plane hysteresis} (a) Hall resistance vs.\ out-of-plane field at 10 K for a CrSBr(30 nm)/FGT(9 nm)/Pt(10 nm) sample. Black dashed lines indicate the estimated coercive fields. b) Positive and negative coercive fields as a function of temperature, together with the average (green), indicating negligible out-of-plane exchange bias.

\newpage
\section{Comparison with antiferromagnetic resonance parameters.}

\begin{figure}[htpb]
    \centering
    \includegraphics[width=0.9\textwidth]{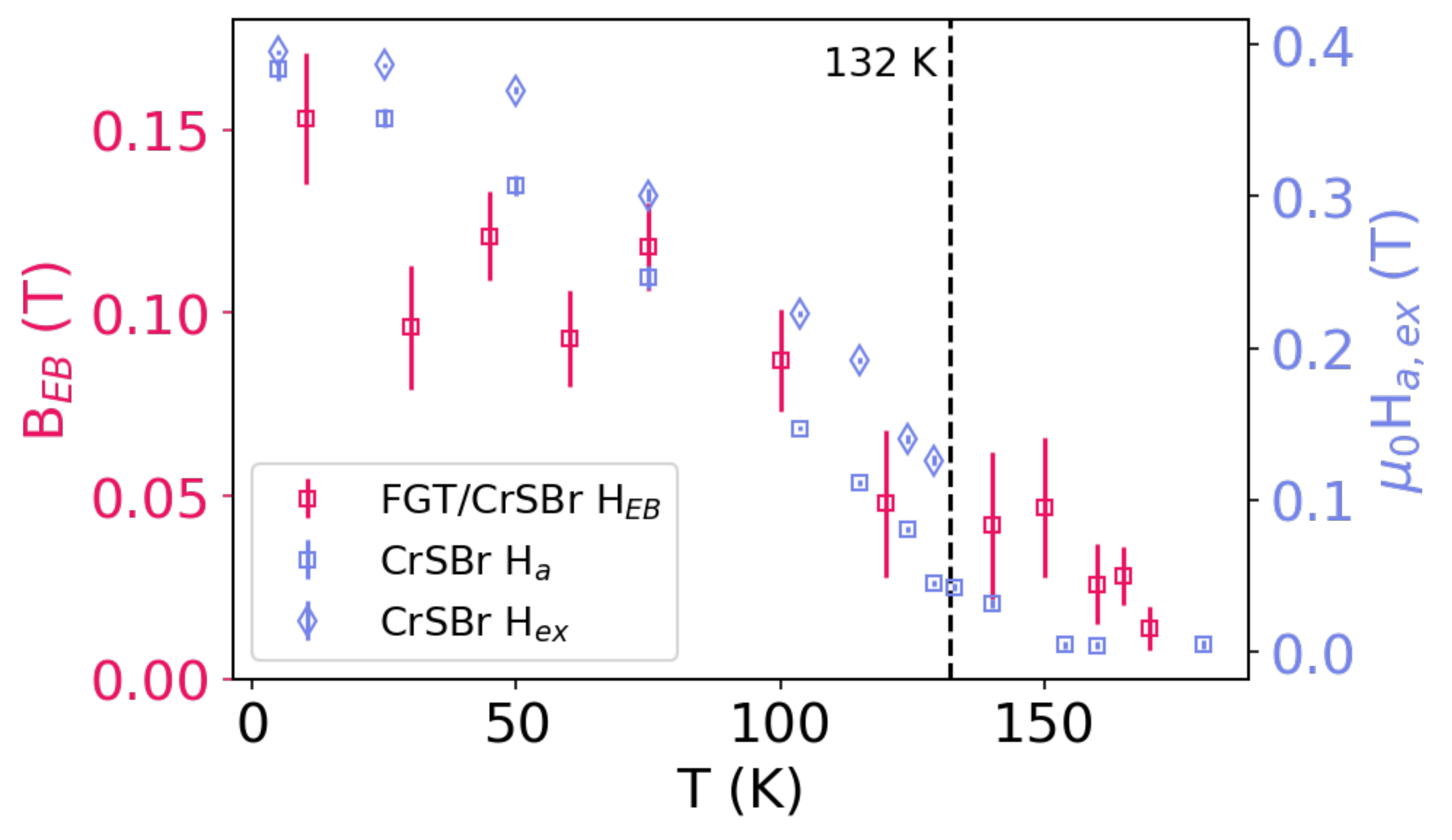}
    \label{SFigure1c}
\end{figure}

\textbf{Fig.~S4. Comparison of the effective exchange bias in a CrSBr(30 nm)/FGT(9 nm)/Pt(10 nm) sample to the anisotropy and exchange parameters in a bulk crystal of FGT.} (red symbols) Effective exchange bias in the CrSBr/FGT/Pt sample, the same data as in Fig.\ 4(e) of the main text.  (blue symbols) Exchange and anisotropy parameters determined from antiferromagnet resonance measurements of a bulk crystal of CrSBr in ref.\ \cite{Cham2022}. 

\newpage
\section{Full temperature series for measurements of pulsed-current switching amplitude versus applied magnetic field.}

\begin{figure}[htpb]
    \centering
    \includegraphics[width=1.0\textwidth]{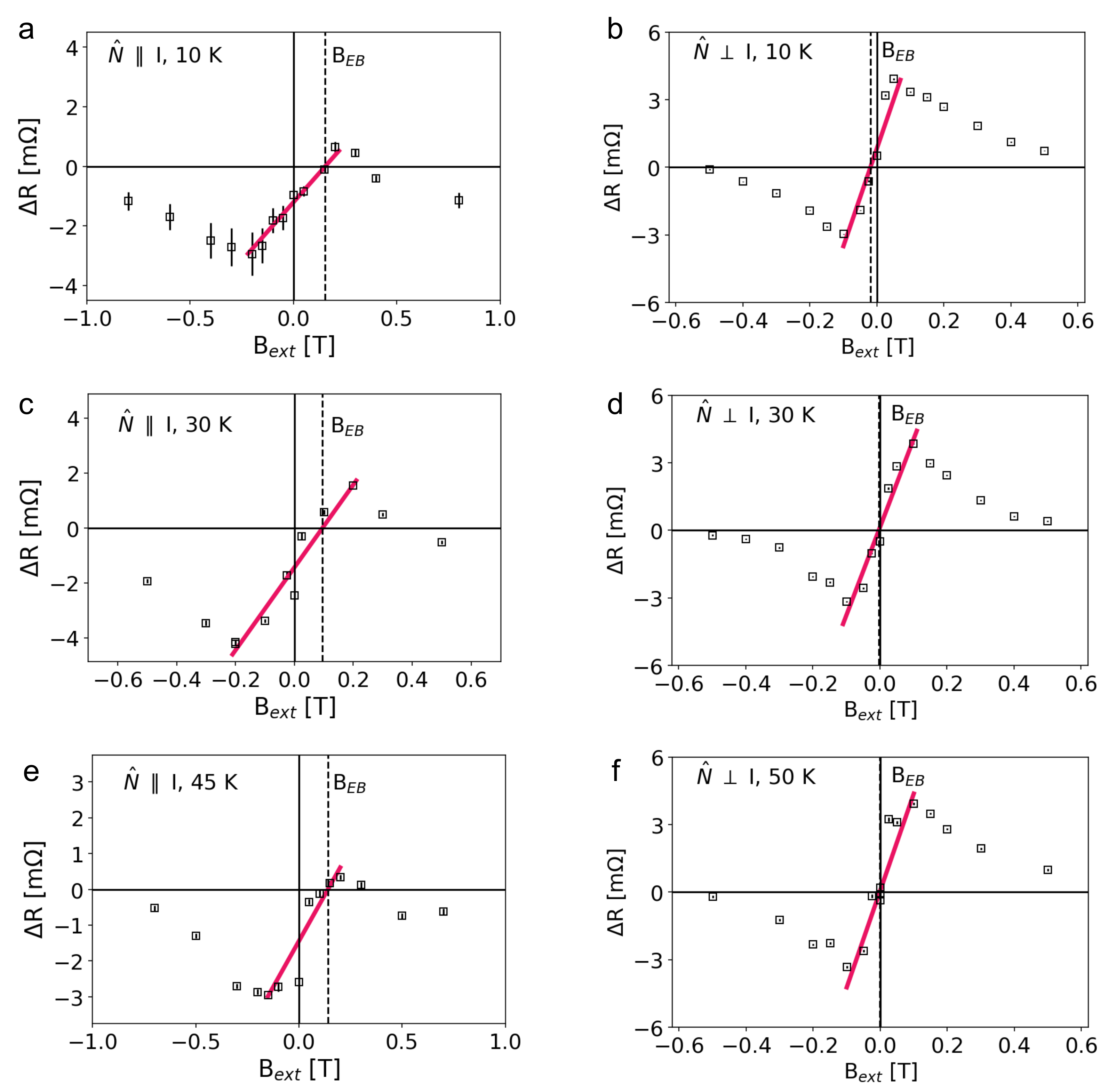}
    \label{SFigure2a}
\end{figure}

\newpage
\begin{figure}[htpb]
    \centering
    \includegraphics[width=1.0\textwidth]{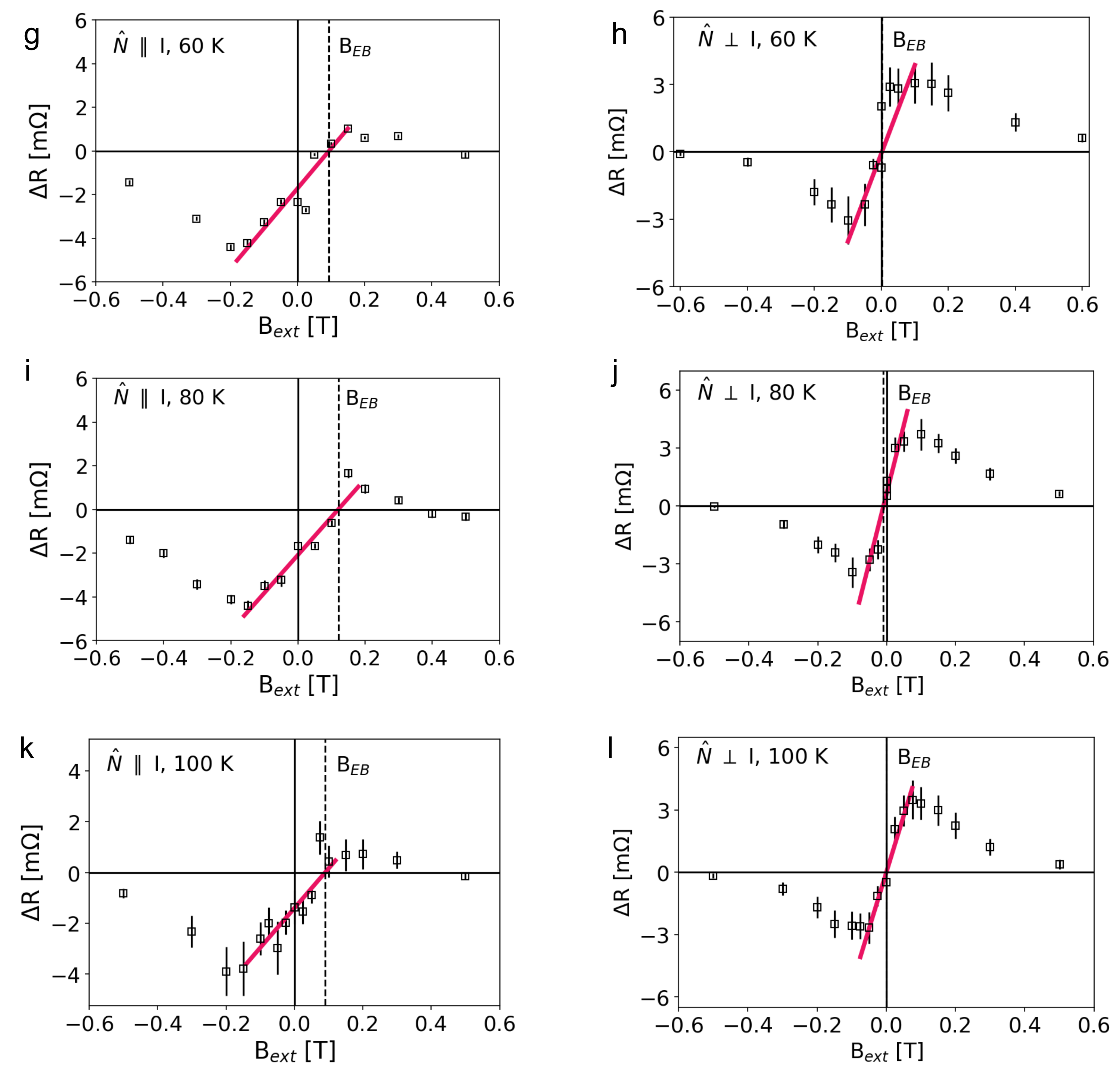}
    \label{SFigure2b}
\end{figure}

\newpage
\begin{figure}[htpb]
    \centering
    \includegraphics[width=1.0\textwidth]{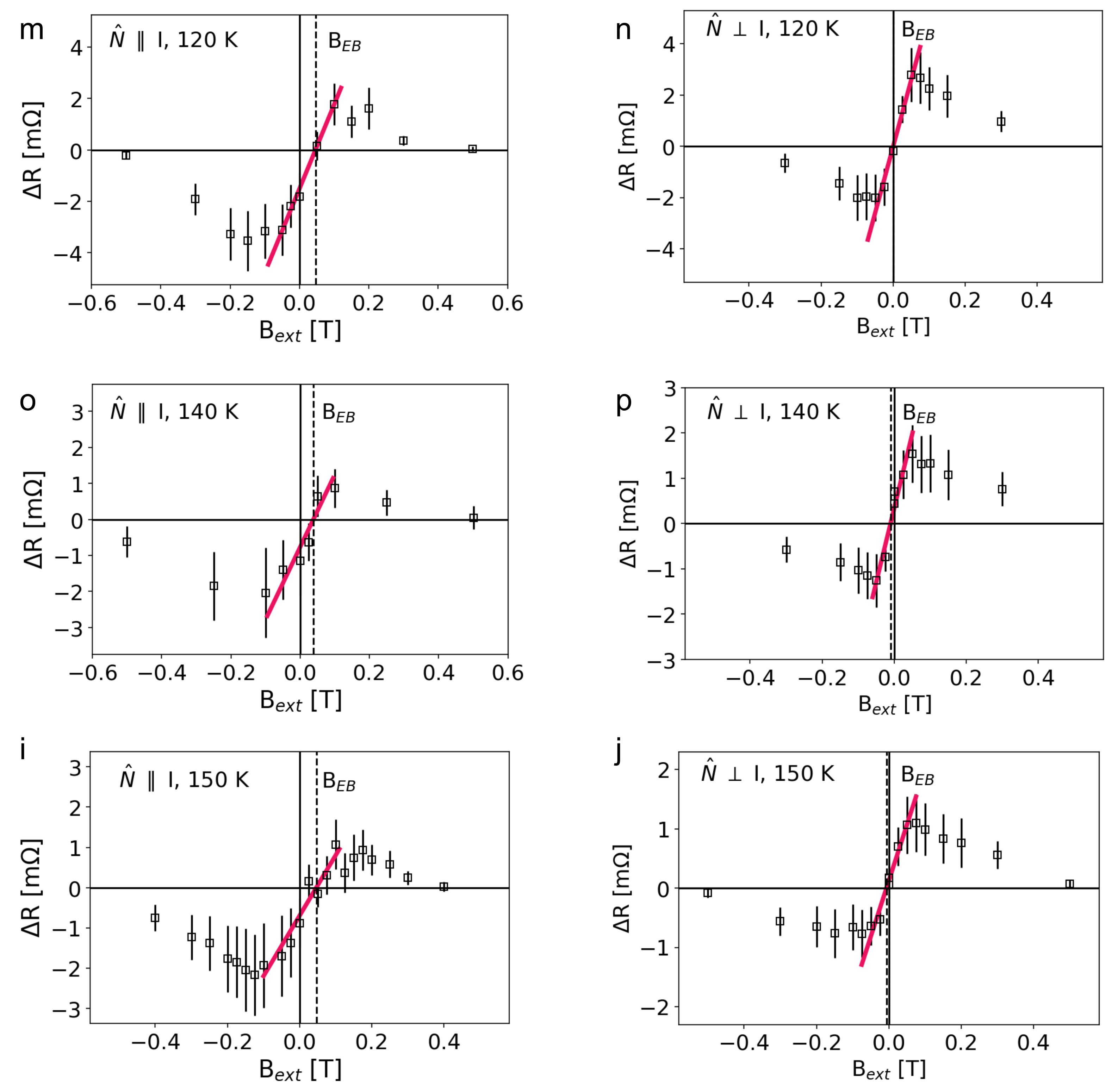}
    \label{SFigure2c}
\end{figure}

\newpage
\begin{figure}[htpb]
    \centering
    \includegraphics[width=1.0\textwidth]{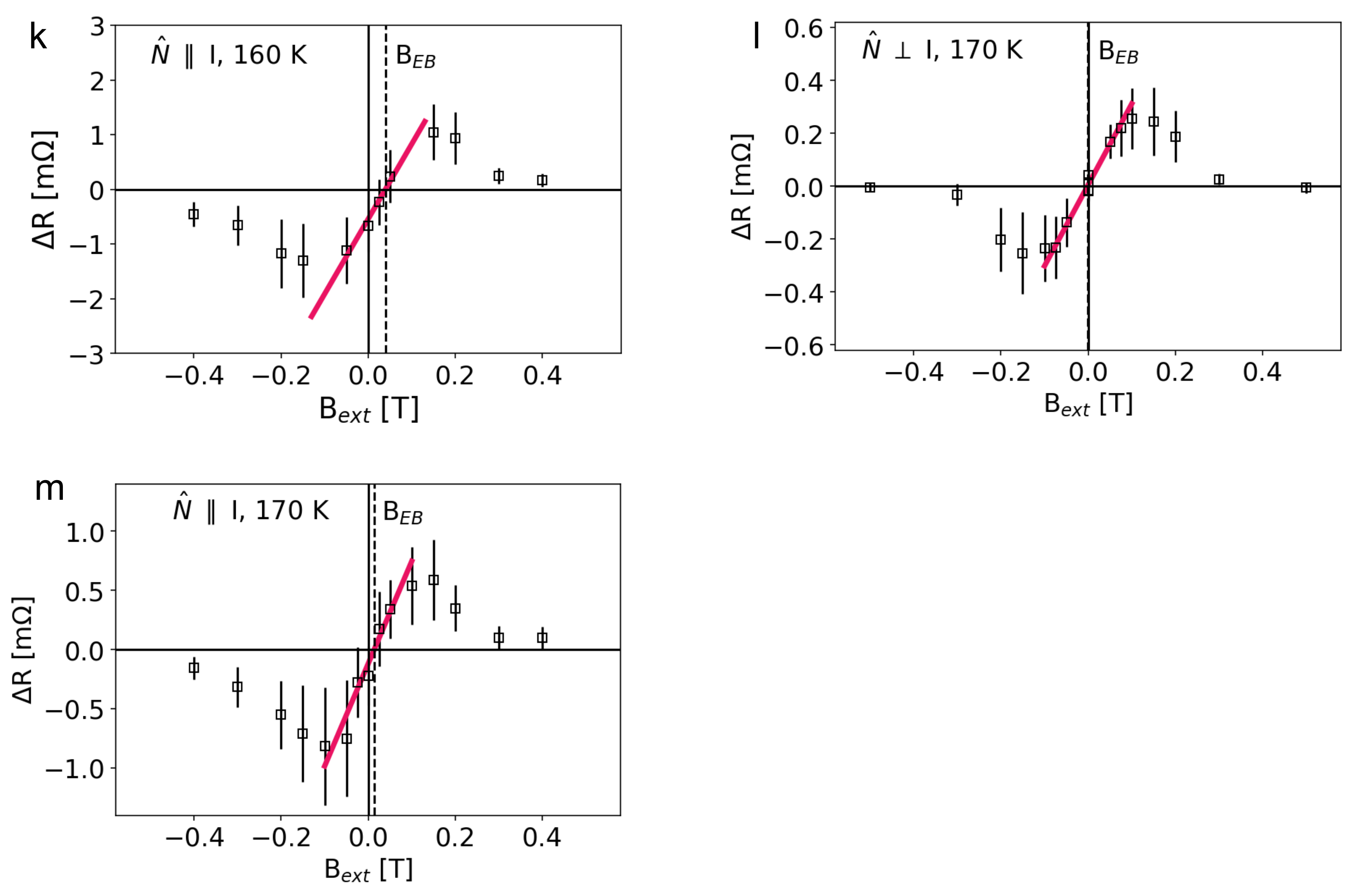}
    \label{SFigure2d}
\end{figure}

\textbf{Fig.~S5. Full temperature series of $\Delta$R vs. B$_\text{ext}$ for the $\hat{\text{N}} \parallel$ I and $\hat{\text{N}} \perp$ I devices described in the main text.} Linear fits (red) to the low field regime are used to estimate the exchange bias field strength, B$_{\text{EB}}$. Dashed lines show B$_{\text{EB}}$ values extracted from the fits. 

\newpage
\section{Switching hysteresis with different C\MakeLowercase{r}SB\MakeLowercase{r} thicknesses}

\begin{figure}[htpb]
    \centering
    \includegraphics[width=0.83\textwidth]{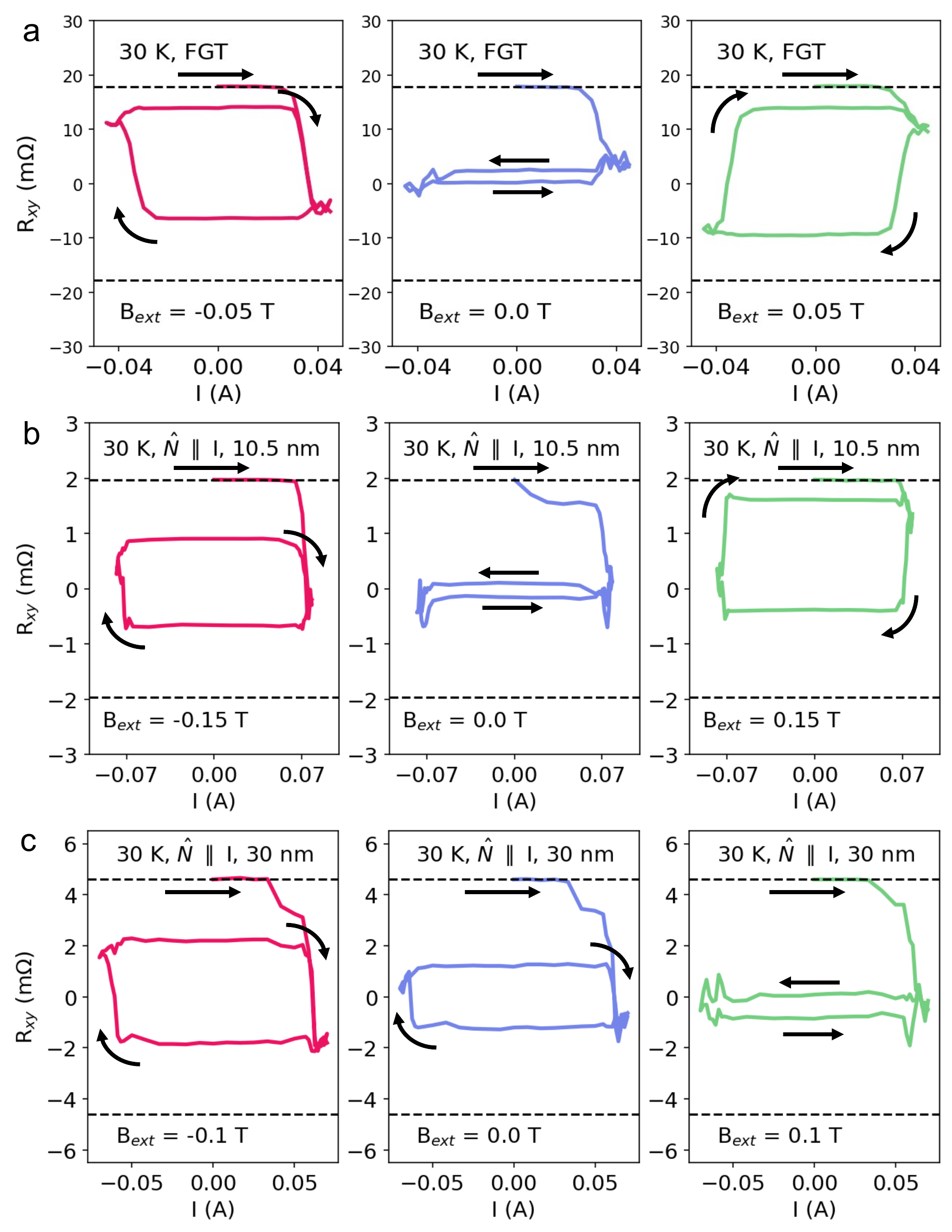}
    \label{SFigure3a}
\end{figure}

\newpage
\begin{figure}[htpb]
    \centering
    \includegraphics[width=0.9\textwidth]{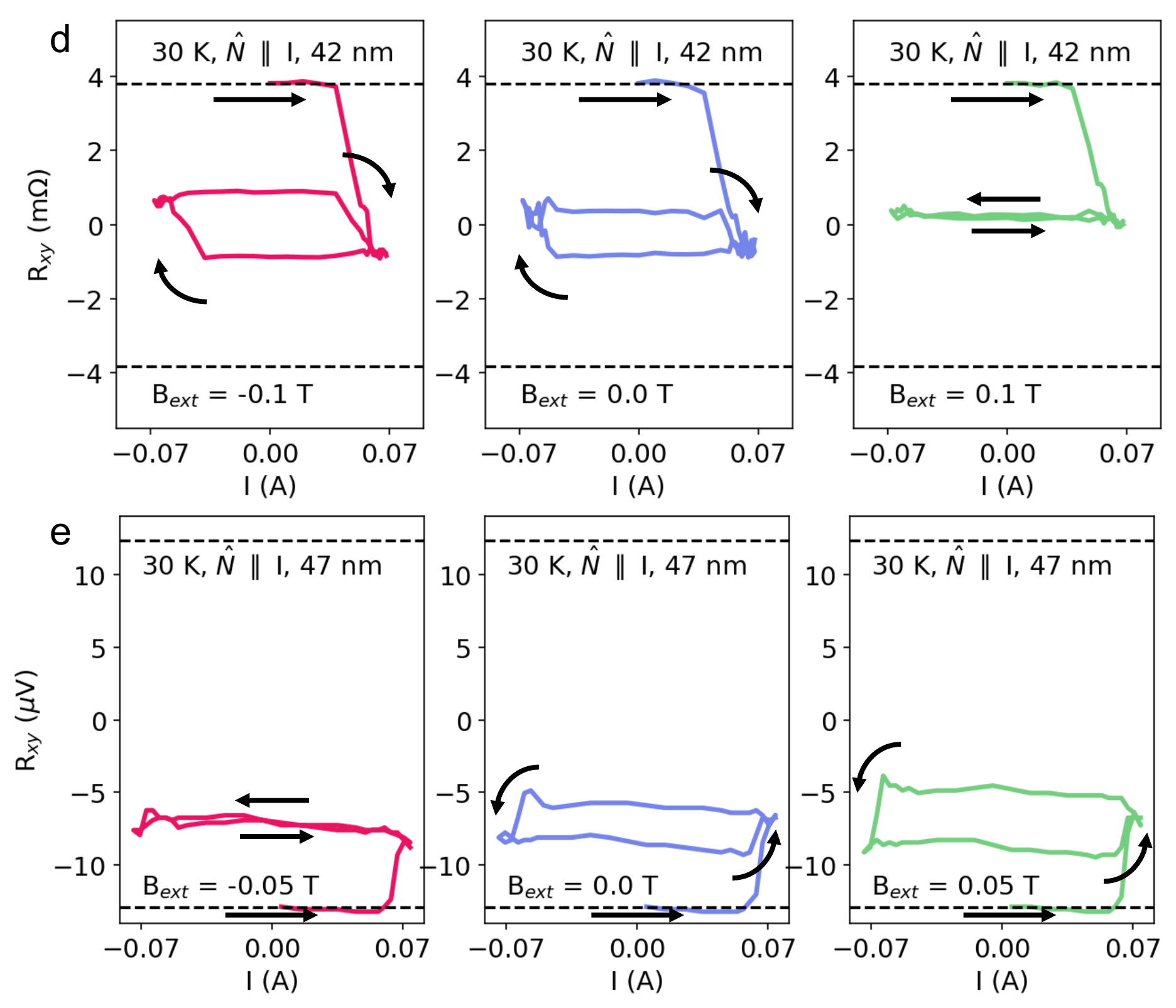}
    \label{SFigure3b}
\end{figure}

\textbf{Fig.~S6. Pulsed-current-switching hysteresis loops of an FGT/Pt sample and CrSBr/FGT/Pt samples with varying thicknesses of CrSBr and varying values of magnetic field applied parallel to the current direction.}  The devices have layer thicknesses (a) FGT(10 nm)/Pt(10 nm) with no CrSBr, (b) CrSBr(10.5 nm)/FGT(9 nm)/Pt(10 nm), (c) CrSBr(30 nm)/FGT(9 nm)/Pt(10 nm), (d) CrSBr(42 nm)/FGT(11.6 nm)/Pt(10 nm) and (e) CrSBr(47 nm)/FGT(9.6 nm)/Pt(10 nm).

\newpage
\begin{figure}[htpb]
    \centering
    \includegraphics[width=0.95\textwidth]{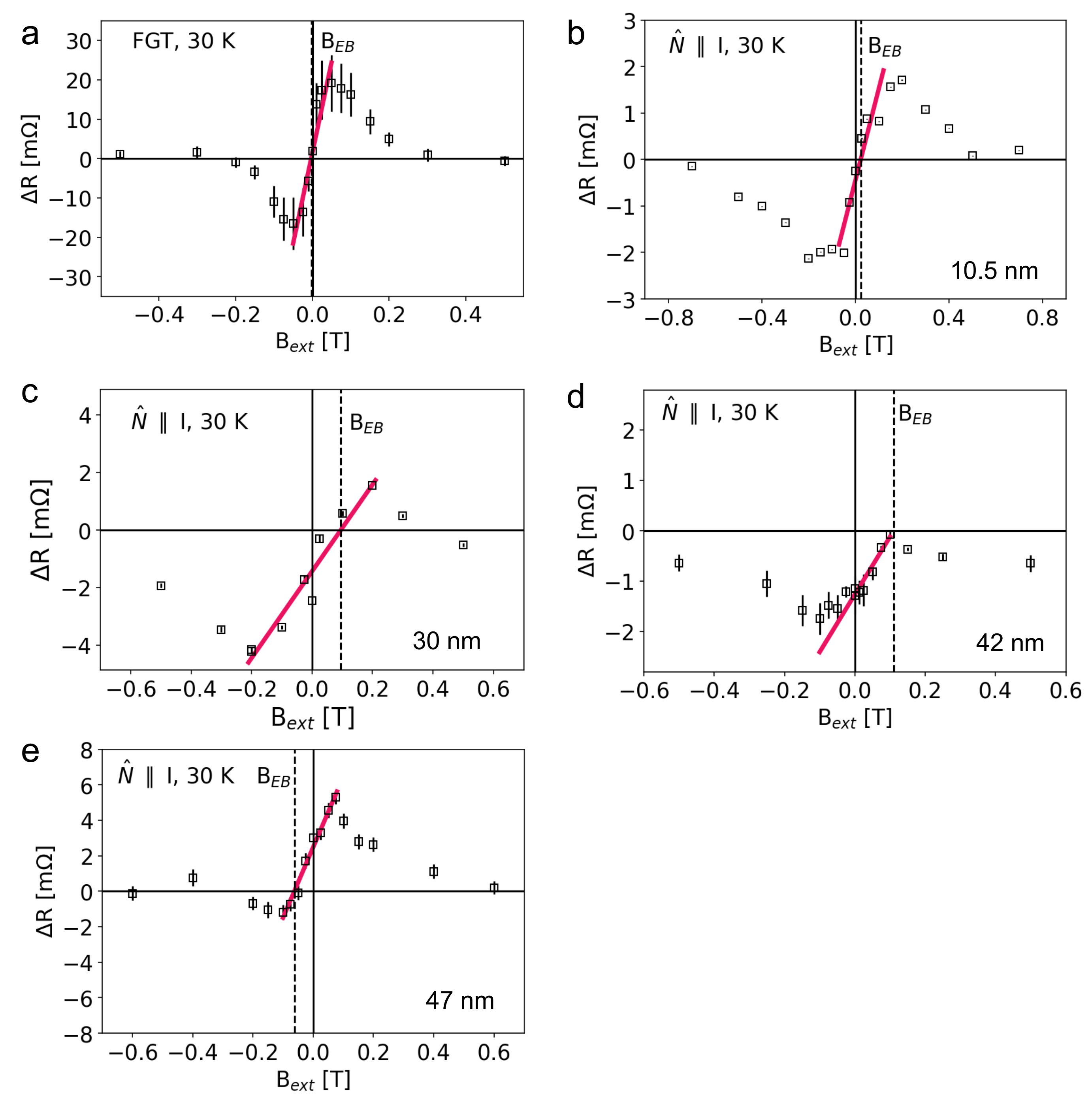}
    \label{SFigure3c}
\end{figure}

\textbf{Fig.~S7. Pulsed-current-switching amplitudes of the devices with different CrSBr thicknesses.}  

\newpage
\section{Field Cooling}
We have investigated the effects of field-cooling on the exchange bias effect observed in CrSBr/FGT/Pt heterostructures, by applying an in-plane magnetic field along the current direction while cooling the sample from  room temperature to 30 K. Pulse-current-switching measurements were then done for cooling fields of 0 T (Fig.~S8a), +8 T (Fig.~S8b) and -8 T (Fig.~S8c). We observe no significant difference in the $\Delta$R vs.\ B$_\text{ext}$ plots, and extract very similar exchange bias field strengths around 0.06 T for all three cases.

\begin{figure}[htpb]
    \centering
    \includegraphics[width=1.0\textwidth]{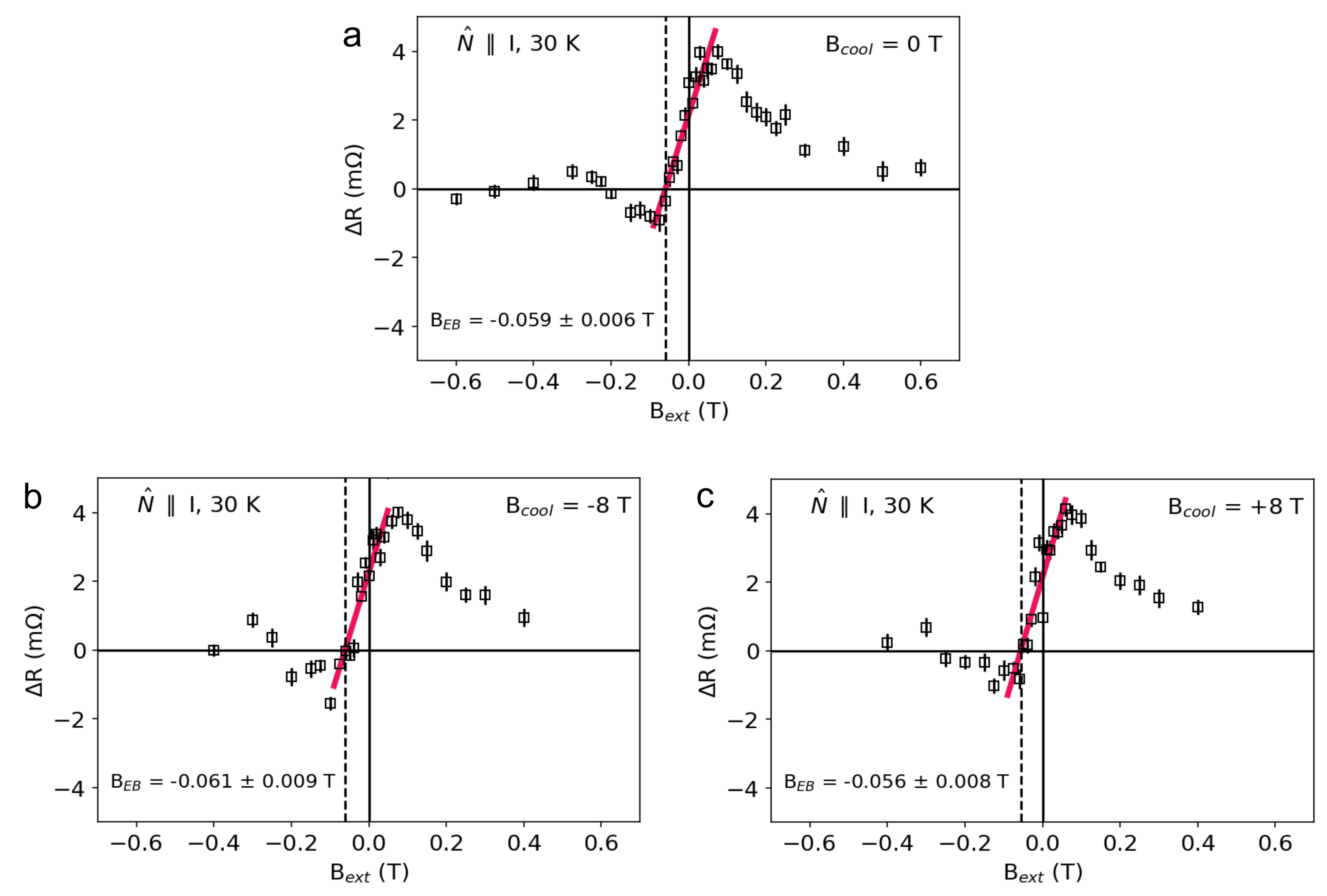}
    \label{SFigure4}
\end{figure}

\textbf{Fig.~S8. Switching amplitude vs.\ external fields with and without field cooling for the CrSBr(47 nm)/FGT(9.6 nm)/Pt(10 nm) device.}

\newpage
\section{Data from an additional device with $\hat{\text{N}} \parallel$ I}
Here we show measurements from a CrSBr(47 nm)/FGT(9.6 nm)/Pt(10 nm) device which support the conclusions in the main text. At 30 K, we observe deterministic partial switching without any external field, and an asymmetry in the switching amplitudes for B$_\text{ext}$ = -0.75 T and +0.75 T (Fig.~S9a). At 160 K, the deterministic switching disappears without an applied field, and the switching chirality is reversed for applied fields of opposite signs (Fig.~S9b). The temperature dependence of B$_\text{EB}$ extracted from the $\Delta$R vs.\ B$_\text{ext}$ plots (Fig.~S9c) show a monotonic decrease with increasing temperature, in qualitative agreement with the device shown in Fig.\ 4 of the main text. 

\begin{figure}[htpb]
    \centering
    \includegraphics[width=0.9\textwidth]{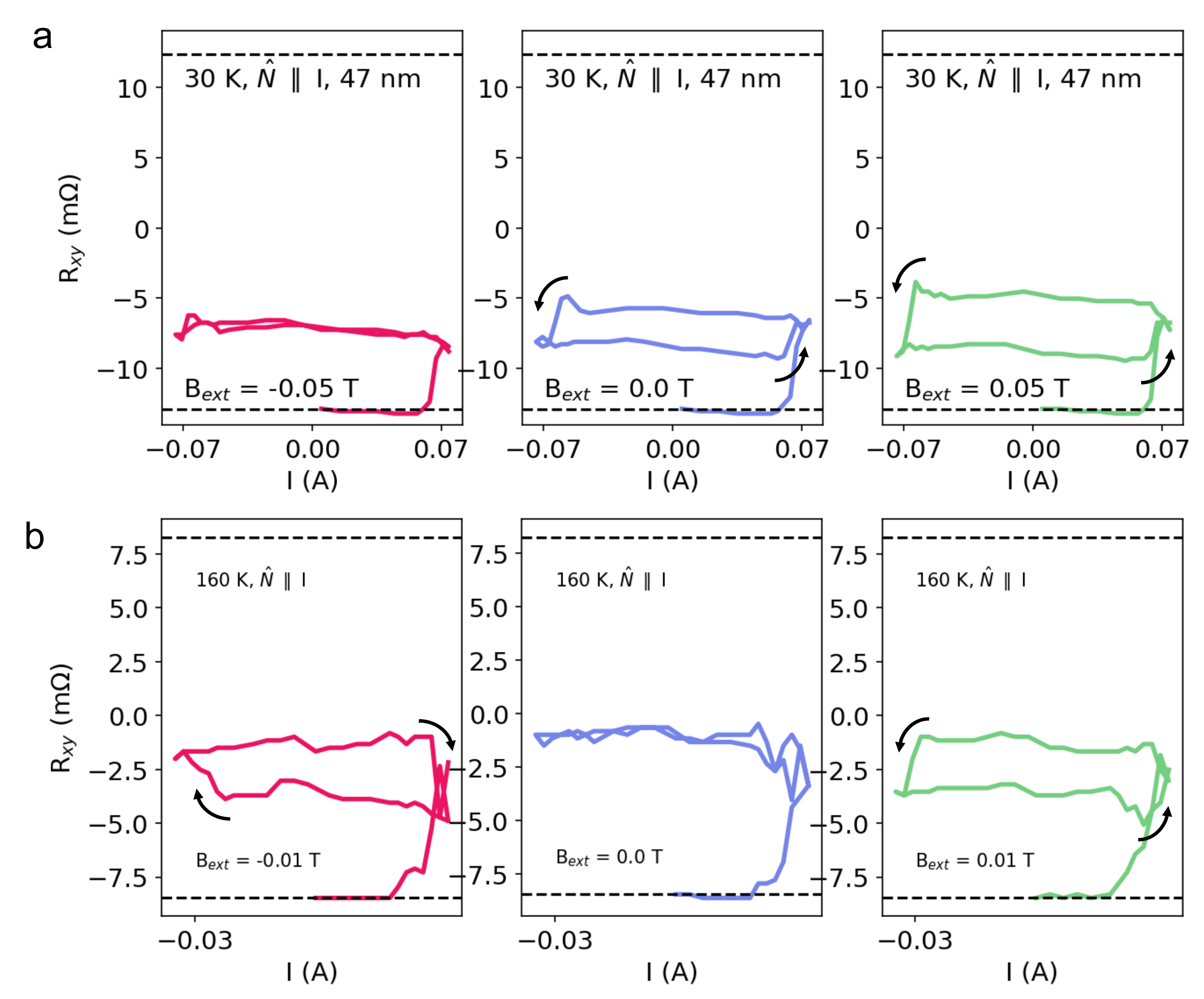}
    \label{SFigure5a}
\end{figure}

\newpage
\begin{figure}[htpb]
    \centering
    \includegraphics[width=1.0\textwidth]{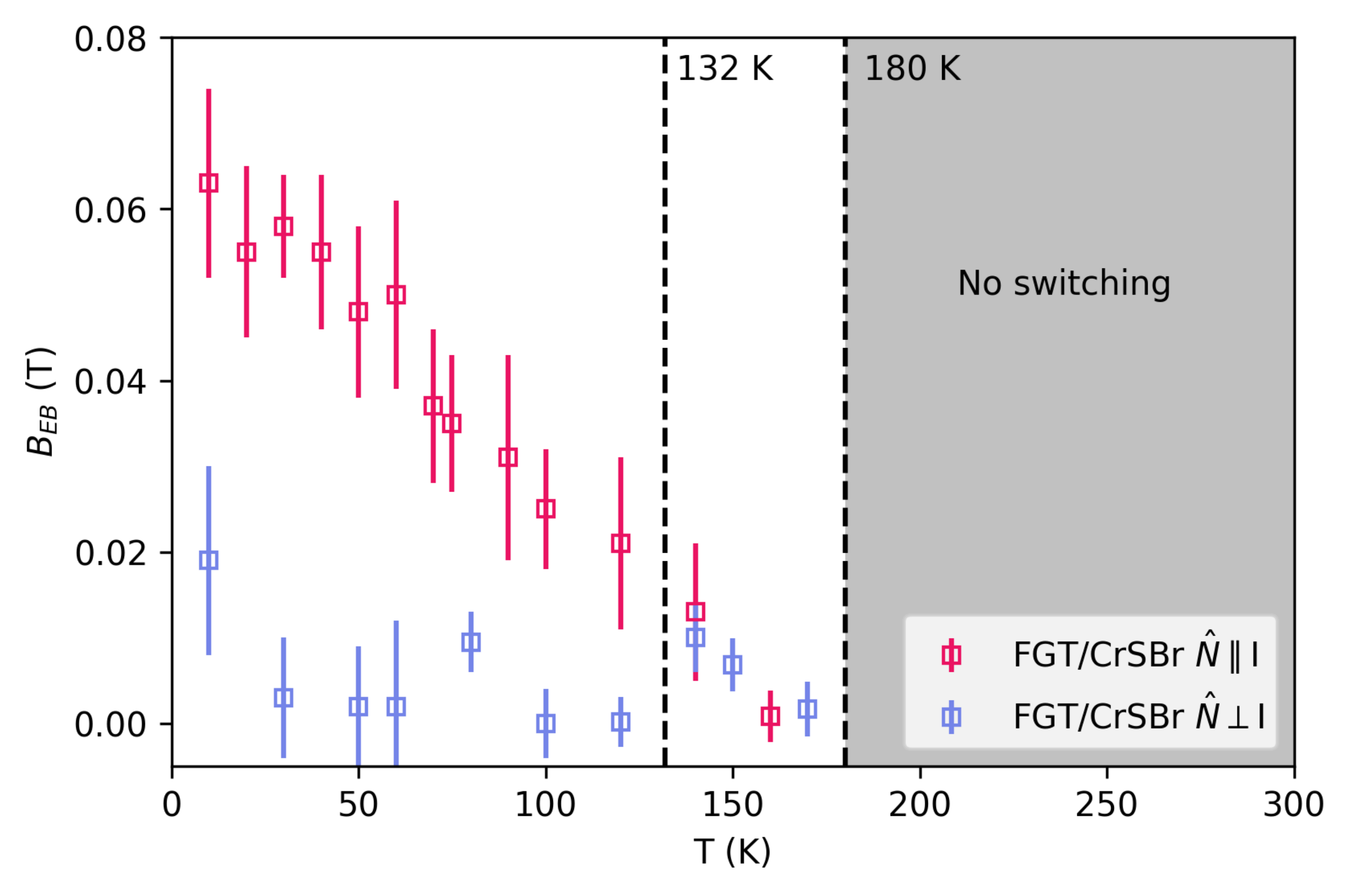}
    \label{SFigure5a}
\end{figure}

\textbf{Fig.~S9. Switching characteristics of a CrSBr(47 nm)/FGT(9.6 nm)/Pt(10 nm) device with $\hat{N} \parallel $ I.} (a,b) Pulse-current-switching hysteresis loops at (a) 30 K  and (b) 160 K. (c) Temperature dependence of the estimated exchange bias B$_{EB}$. 

\newpage
\section{Initialization of the FGT magnetization at positive and negative values}
We show pulse-current-switching hysteresis loops of a CrSBr/FGT/Pt device with $\hat{N} \parallel$ I where the FGT was first initialized in the M$_\text{z}$+ state (Fig.~S10a) or the M$_\text{z}$- state (Fig.~S10b). In both cases, we observe a the same deterministic switching at 0 T and asymmetric switching amplitudes for positive and negative applied fields. This suggests that the initialization process has a negligible effect on the exchange-bias field which mediates the field-free deterministic magnetic switching.

\begin{figure}[htpb]
    \centering
    \includegraphics[width=0.9\textwidth]{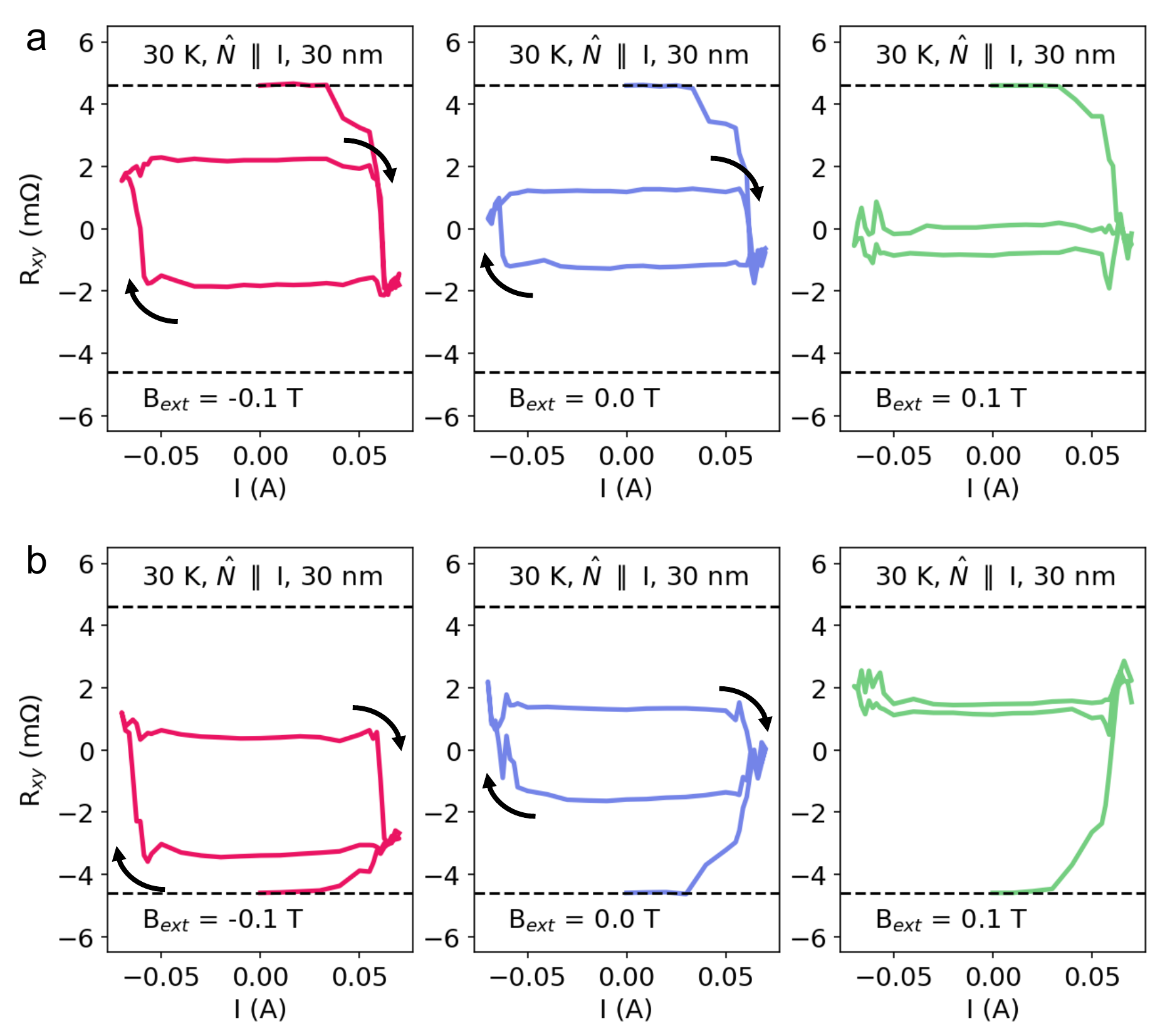}
    \label{SFigure6}
\end{figure}

\textbf{Fig.~S10. Comparison between current-driven hysteresis loops for the CrSBr(30 nm)/FGT(9 nm)/Pt(10 nm) sample initialized in the M$_\text{z+}$ and M$_\text{z-}$ states.}

\newpage
\section{Micromagnetic Simulations of current-induced switching}

In order to gain insights into the role of thermal effects on the incomplete magnetization reversal driven by the spin-orbit torque,  we conducted micromagnetic simulations of the field dependence of $\Delta R$ using \textit{mumax}${}^3$, an open-source GPU-accelerated micro-magnetic simulation program\textsuperscript{\cite{Vansteenkiste2014}}. We first simulated simple FGT/Pt devices without any exchange bias. The simulated system consists of a 200 nm $\times$ 200 nm $\times$ 4 nm ferromagnetic slab, discretized into 256 $\times$ 256 cells in the xy-plane. These dimensions were chosen both for manageable simulation speeds as well as to ensure that the dimensions of each cell are $\leq 3/4\;\times$ the exchange length of the system. Defects were added to the system geometry to reduce domain mobility: a randomly-generated grain structure was imposed on the system via the Voronoi tessellation method, and small triangular notches were subtracted from the system edges to simulate edge defects (Fig.\ S11). Each grain has a 5 nm size scale, slight variation in uniaxial anisotropy strength ($\leq$ 20\%) and anisotropy direction ($\leq$ 10$^{\circ}$), and the exchange coupling between neighboring grains was reduced by 10\%. The magnetic material parameters were chosen to emulate the hard magnetic properties of FGT\textsuperscript{\cite{Wang2020}} -- a saturation magnetization of $3.8\times 10^5$ A/m, exchange stiffness of $1 - 4$ pJ/m, out-of-plane uniaxial anisotropy of $100-150\times 10^4$ J/m${}^3$ and damping factor of $0.02$. While its origin is still unclear, a bulk DMI interaction of $0.5-1$ mJ/m${}^2$ was included as motivated by the literature\textsuperscript{\cite{Meijer2020}}.

\vspace{0.5cm}
We simulate the effects of current pulses via a damping-like spin-orbit torque
 \begin{align}
    \vec{\tau}_\text{ST} = \theta_{SH}j_{Pt}(\hat{m}\times(\hat{y}\times\hat{m}))
\end{align}
where $\theta_{SH}$ is the spin-Hall ratio, $j_{Pt}$ is the charge current density in the Pt layer, $\hat{m}$ is the unit vector of the local FGT magnetization, and $\hat{y}$ is a unit vector in the sample plane perpendicular to the applied current.  This torque reorients the local magnetization in the FGT via the Landau-Lifschitz-Gilbert equation
\begin{align}
    \frac{\partial \vec{m}}{\partial t} = \vec{\tau}_\text{LL} + \vec{\tau}_\text{ST}
\end{align}
where $\vec{\tau}_\text{LL}$ accounts for the torques due to the applied magnetic field, the anisotropy field, DMI, and damping. We take into account a symmetry-breaking magnetic field applied along the current direction. 

\vspace{0.5cm}
The actual simulations consist of initializing the system in a +$z$ saturated state at 0 K and with a magnetic field applied along $x$. 
Current pulses were applied starting from 0 mA up to 15 mA, down to -15 mA, then back again to 15 mA in steps of about 1.5 mA.
Building on the suggestion by X.\ Wang et al.\ that the partial switching behavior may be due to a Joule heating effect\textsuperscript{\cite{Wang2019}}, we assume that current pulses (in addition to applying spin-orbit torque) produce a temperature increase approximately proportional to the current magnitude (e.g., a temperature rise of 85 K under a current pulse of 15 mA), which is incorporated into the simulation by scaling the magnetic material parameters of the FGT based on temperature-dependence data in the literature\textsuperscript{\cite{Vertesy2003}}, scaled down for the lower FGT Curie temperature of $\sim$200 K. The temperature dependencies assumed are plotted in Fig.\ S12), alongside the resulting magnetization states when temperature is raised from an initial out-of-plane saturated state. Similar to observations reported in the literature\textsuperscript{\cite{Meijer2020}}, as the temperature nears T${_c}$ the FGT displays a pattern of winding, spiral domains. The formation of these domains places the T${}_c$ of the simulated system within 150-200 K. It was also found that the thermally-formed domains would quickly collapse if the temperature were to abruptly fall back to 0 K under a step-function approximation; therefore, the 2 ns current/temperature pulses were followed by a 4 ns period in which both the temperature and spin-torque fall off exponentially. At the end of each 6 ns pulse-and-rest period, the system's average magnetization along $z$ was saved along with a color-plot image of the corresponding domain structure. 


\vspace{0.5cm}
For simplicity we model temperature only by the scaling of the magnetic parameters, excluding any stochastic thermal field. We find that if we start the system in a saturated $+z$ magnetization state and apply successive current pulses, the system undergoes a partial magnetization reversal and displays deterministic switching between two intermediate spatially non-uniform states (Fig.~S13a). The weakening of the magnetic anisotropy caused by heating allows for randomization of the magnet's domain structure at high current. However, the influence of the spin-orbit torque results in a systematic preference for domains of one parity over the other. Upon cooling, we are left with a mixed-domain state with domain walls pinned by defects. As the current pulses increase in the opposite direction,  domain walls are shifted until the competing effects of thermal randomization, spin torque, and defect pinning determine a second stable configuration. The resulting hysteresis loop corresponds to this back-and-forth redistribution of domain walls between metastable states. 

\vspace{0.5cm}
The simulated switching amplitude $\Delta R$ in the absence of any exchange bias (Fig.~S13b) is seen to follow an anti-symmetric dependence on magnetic field similar to the results of our pulse-current switching measurements of FGT/Pt (Supplementary Material Section VI, Fig.\ S6a and Fig.\ S7a) as well as CrSBr/FGT/Pt in the absence of a symmetry breaking exchange bias field (Main text figures 4a-c). These simulations suggest that Joule heating and domain formation in FGT may account for the both the incomplete switching characteristic and the approximately anti-symmetric Gaussian field dependence.

\vspace{0.5cm}
A simple model of the exchange bias due to a capping layer of CrSBr was then tested. The 200 nm $\times$ 200 nm $\times$ 4 nm slab was divided into two equally-sized layers, defined as distinct material regions in \textit{mumax}${}^3$. The effective separation between cells in the $z$-direction was thus 2 nm, over double the separation between cells within the xy-plane. This was to simulate the weakened inter-layer interaction of a van der Waals magnetic material. The top layer was subjected to an additional external field of 0.25 T along the $+x$-direction, modelling the effective exchange field felt by the upper layers of the FGT due to the antiferromagnetic material above it. The same current-induced switching simulation was then run using this modified system. This modified system displays deterministic switching at zero applied field (Fig.~S13c).
The corresponding field dependence of the switching amplitudes (Fig.~S13d) exhibits a non-zero value at 0 T ($\Delta R$ $>$ 0), and a quenching of the switching ($\Delta R$ = 0) at a non-zero negative field similar to the experimental results. Notably, the exchange bias is not equivalent to an applied magnetic field, in that the exchange bias does not simply shift the overall $\Delta R$ vs.\ B$_\text{ext}$ curve horizontally. Instead, the functional form of the $\Delta R$ vs.\ B$_\text{ext}$ curve is also changed, possessing an asymmetry in the hysteresis magnitudes for positive and negative B$_\text{ext}$.  This asymmetry is not as pronounced in the simulation as in some of our measurements (Fig.~4d), but this result nevertheless indicates that exchange bias is capable of generating the main qualitative features by which the switching in our CrSBr/FGT/Pt samples differs from just FGT/Pt.

\vspace{1.0cm}
\begin{figure}[htpb]
    \centering
    \includegraphics[width=1.0\textwidth]{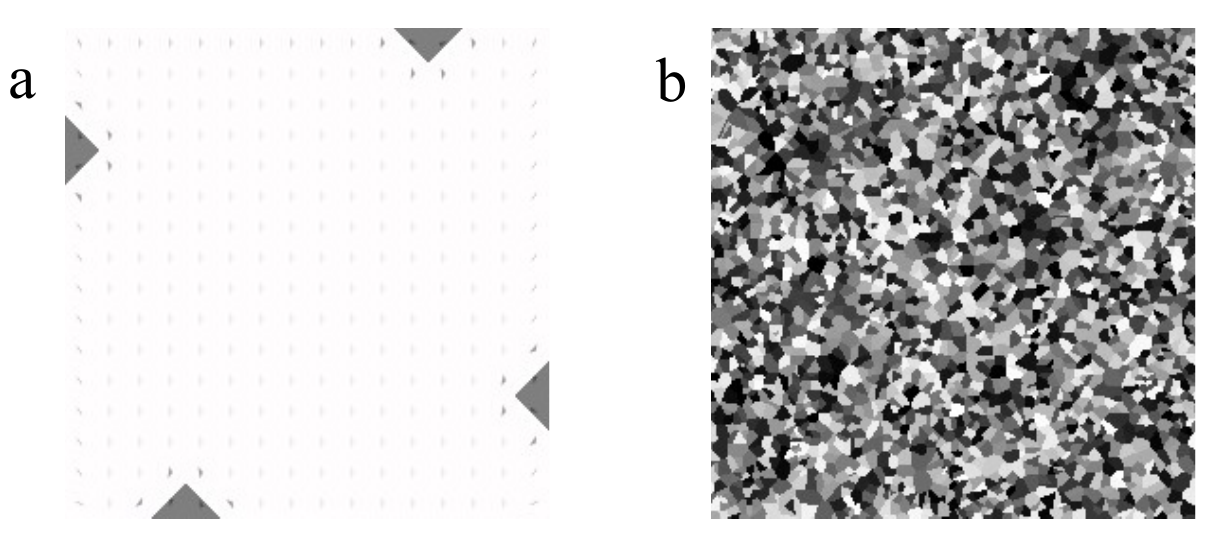}
    \label{SimFigure1}
\end{figure}
\vspace{0.5cm}
\textbf{Fig.~S11. Material defects incorporated into the simulated FGT system.} (a) Triangular notches subtracted from the sides of the simulation window emulate structural defects that contribute to domain-wall pinning and symmetry-breaking in the material. (b) A randomly-generated grain structure is imposed on the system that reduces domain-wall mobility and homogeneity of the sample.

\newpage
\begin{figure}[htpb]
    \centering
    \includegraphics[width=0.8\textwidth]{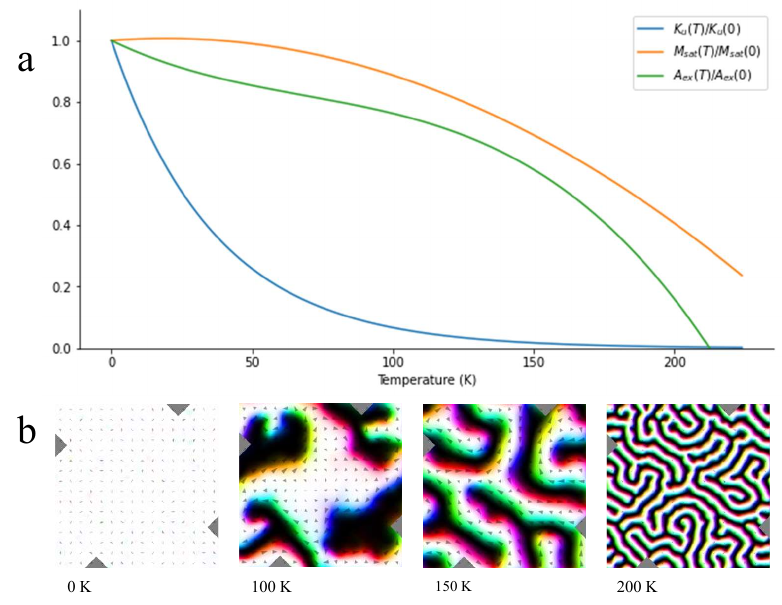}
    \label{Figure5}
\end{figure}
\textbf{Fig.~S12. Temperature-dependence model used in the micromagnetic simulations.} (a) Plotted are the uniaxial anisotropy (blue), saturation magnetization (orange) and exchange stiffness (green) as functions of temperature, normalized by their values at 0 K. (b) Snapshots of domain structure as developed in a temperature-sweep simulation at 0 K, 100 K, 150 K, and 200 K.

\newpage
\begin{figure}[htpb]
    \centering
    \includegraphics[width=1.0\textwidth]{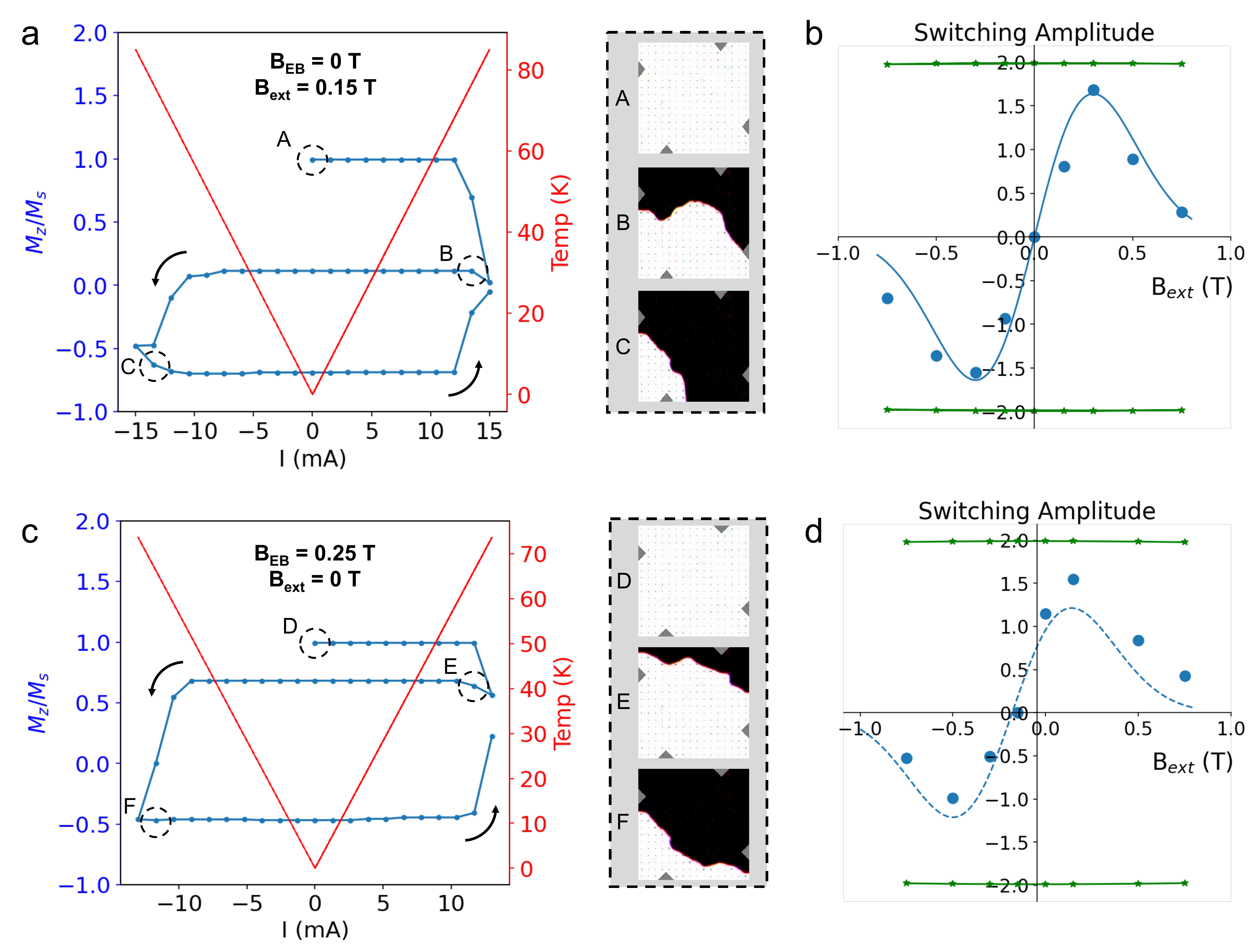}
    \label{Figure6}
\end{figure}
\textbf{Fig.~S13. Simulations of deterministic current-induced switching.} (a) (blue curve) Simulated hysteresis loop for pulsed-current  magnetization switching of an FGT layer without exchange bias, for an external field of 0.15 T applied along the current direction.  (red curve) Assumed current-induced increase in sample temperature, following ref.\ \cite{Wang2019}. The domain patterns shown at right correspond to a 200 nm $\times$ 200 nm area for (A) the initial fully-saturated state, (B) the non-uniform state at zero current after pulsing the current to $+$12 mA, and (C) the non-uniform state at zero current after pulsing the current to $-$13 mA. (b) (points) Simulated switching amplitudes as a function of applied magnetic field. (blue curve) Antisymmetric Gaussian fit serving as a guide to the eye. (green points) Out-of-plane saturated magnetization for each value of the applied field. (c) Simulated hysteresis loop for a toy model with an exchange bias of 0.25 T and no applied magnetic field. The domain patterns shown at right correspond to (D) the initial configuration and (E,F) non-uniform configurations at zero current after pulsing to positive and negative currents. (d) (points) Field dependence of the switching amplitude for the exchange-biased toy model.  The dashed blue curve is a horizontally-shifted antisymmetric Gaussian guide to the eye, to illustrate asymmetry in the simulated switching amplitudes.
 
\section{Optical image of control device with C\MakeLowercase{r}SB\MakeLowercase{r} layer oriented such that $\hat{N} \perp I$}
\begin{figure}[htpb]
    \centering
    \includegraphics[width=0.45\textwidth]{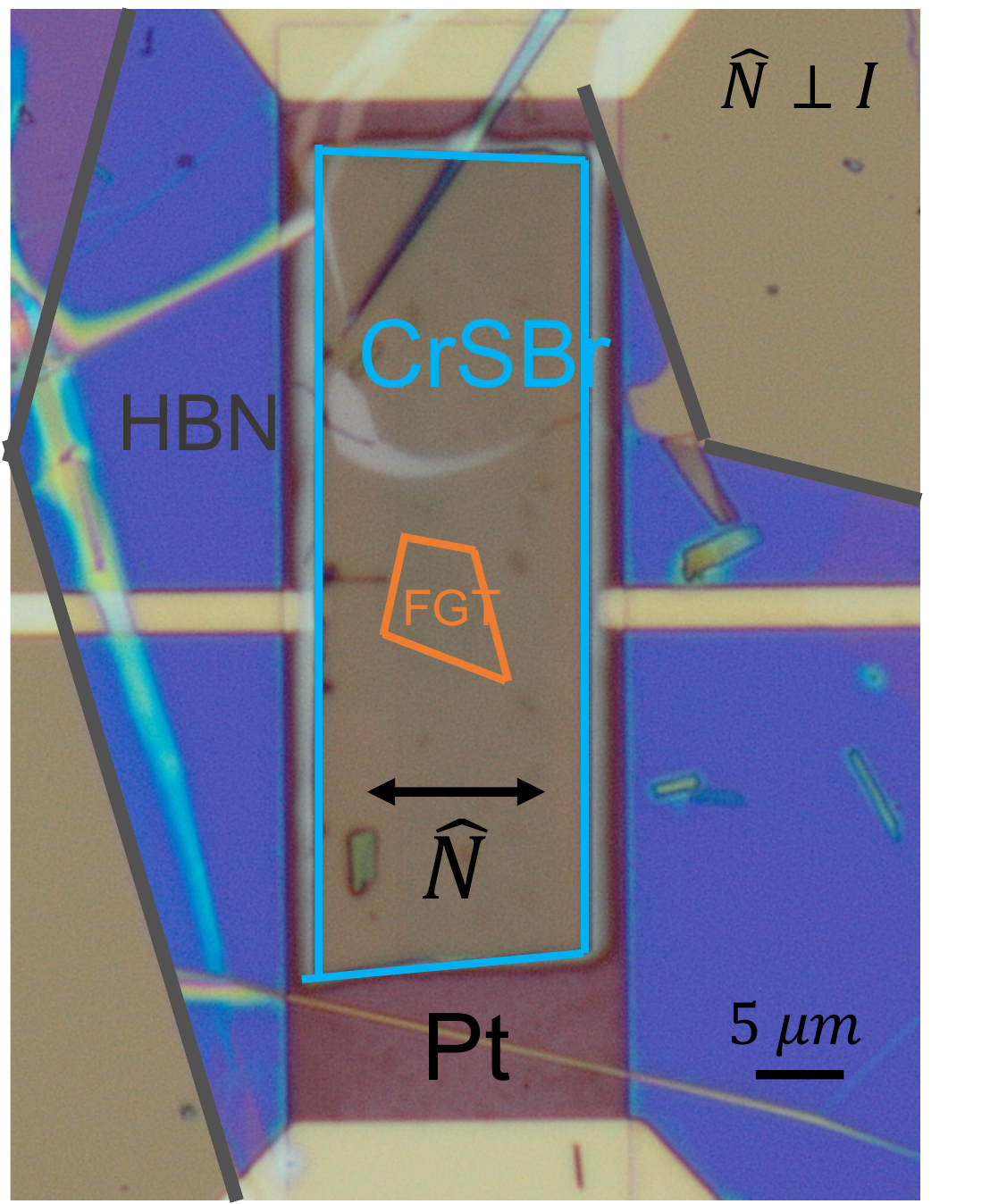}
    \label{SFigure1a}
\end{figure}

\textbf{Fig.~S14. Optical image of device with CrSBr $\hat{N} \perp I$.} CrSBr(37 nm)/FGT(12 nm)/Pt(10 nm) device with short edge of CrSBr perpendicular to the current direction. In this configuration, the exchange bias does not provide a symmetry-breaking field along the current direction for field-free SOT switching.

\end{document}